\documentclass[preprint]{elsarticle}

\usepackage[utf8]{inputenc}
\usepackage[UKenglish]{babel}

\usepackage{subfiles}
\usepackage{graphicx}
\usepackage{enumitem}

\usepackage{amsmath}
\usepackage{amssymb} 
\usepackage{mathtools}
\usepackage{units}
\usepackage{url}
\usepackage{dsfont} 

\usepackage{amsthm} 
\newtheorem{satz}{Satz}

\newtheorem{proposition}[satz]{Proposition}

\newtheorem{theorem}[satz]{Theorem}
\newtheorem{remark}[satz]{Remark}
\newtheorem{example}[satz]{Example}

\usepackage{glossaries}
\newacronym{ice}{ICE}{integral constitutive equation}
\newacronym{itt}{ITT}{integration through transients}
\newacronym{mct}{MCT}{mode-coupling theory of the glass transition}
\newacronym{ode}{ODE}{ordinary differential equation}
\newacronym{fve}{FVE}{fractional viscoelasticity}
\newacronym{srr}{SRR}{strain-rate relaxation}
\newacronym{so}{SO}{stress-overshoot}
\newacronym{avie}{AVIE}{autoconvolution Volterra integral equation}

\usepackage{orcidlink}

\begin{document}


\begin{frontmatter}

\title{Nonlinear integral equations for stress-controlled rheology: Collocation methods for Volterra equations with functional kernels}

\author[1,2]{Stephan Domann\orcidlink{0009-0002-1610-398X} \corref{cor1} }
\author[1,2]{Thomas Voigtmann\orcidlink{0000-0002-1261-9295}}

\cortext[cor1]{Corresponding author}
\affiliation[1]{
organization={Institute for Frontier Materials on Earth and in Space, German~Aerospace~Center~(DLR)},
adressline={Linder H\"ohe},
city={K\"oln},
postcode={51147},
country={Germany}
}

\affiliation[2]{
organization={Institut f\"ur Theoretische Physik, Heinrich-Heine-Universit\"at~D\"usseldorf},
adressline={Universit\"atstraße~1},
city={D\"usseldorf},
postcode={40225},
country={Germany}
 }


\begin{abstract}
We present a class of numerical collocation methods
to solve a typical inverse problem
that arises in the stress-controlled rheology of complex fluids with
pronounced history-dependent response, where
the relation between stress and flow rate is given by
an integral constitutive equation as the material law.
We demonstrate the method by calculating the creep response of a nontrivial
model of a shear-thinning viscoelastic fluid where a stress overshoot in
rate-controlled startup flow leads to ``sudden yielding'' in stress-controlled
creep.
Our method is able to address the long-time behavior of the model,
distinguishing the amorphous solid from the stress-induced yielded fluid,
and from the in-between case of infinite deformation that never achieves
a stationary shear rate (proper creep).
To this end, we implement a method with flexible time-stepping, and prove
uniqueness of the solution and
convergence of the methods for a broad class of highly viscoelastic
non-linear rheological laws in integral form.
\end{abstract}

\end{frontmatter}

\section{Introduction}
The description of the flow of non-Newtonian fluids rests on
\emph{constitutive laws:} They link the stress tensor that enters the
coarse-grained macroscopic description based on the Navier-Stokes equations,
with the microscopic material models.
Recall the simplest example of a constitutive law of a fluid,
the Newtonian fluid model: It links the velocity-gradient tensor
$\dot{\boldsymbol\gamma}=\vec\nabla\vec u$ of the fluid velocity field
$\vec u$, linearly to the Euler stress tensor $\boldsymbol\sigma$.
Respecting the epxected symmetries of the stress tensor, one obtains
for incompressible flow, $\vec\nabla\cdot\vec u=0$, the relation
$\boldsymbol\sigma
=\eta(\dot{\boldsymbol\gamma}+\dot{\boldsymbol\gamma}^T)$,
where $\eta$ is the Newtonian viscosity.

Many materials of technological interest go far beyond the simple
Newtonian fluid. The stress-response to a flow can be significantly
delayed in time due to slow internal relaxation of the fluid's microstructure,
and it can be non-linear especially in soft materials where the intrinsic forces
are comparable to deformation forces. In principle, also spatial
non-locality could be considered, but for simplicity we do not deal
with this in the present paper.

A natural form to write a constitutive equation for such a material
is thus an \gls{ice}:
\begin{equation}\label{eq:ice_general}
  \boldsymbol\sigma(t)=\eta(\dot{\boldsymbol\gamma}(t)
  +\dot{\boldsymbol\gamma}^T(t))
  +\int_{t_0}^t {\mathbb G}(t,t',[\dot{\boldsymbol\gamma}]):
  \dot{\boldsymbol\gamma}(t')\,dt'\,,
\end{equation}
if the flow is started at some time $t_0<t$ in the past.
The delayed and potentially non-linear response is encoded in a history integral
whose kernel is a response function $\mathbb G$. It is a fourth-rank
tensor field, obeying certain symmetries that we will not elaborate in
detail here. In the presence of nonlinear response, it has a functional
dependence on the velocity-gradient tensor $\dot{\boldsymbol\gamma}$,
denoted by square brackets.
The first term on the r.h.s.\ of Eq.~\eqref{eq:ice_general}
is an instantaneous Newtonian contribution, representing degrees of freedom
that can be assumed to relax instantaneously on the time scale of change
of the considered flow.
In principle it should be encoded in the integral kernel as a fast-decaying
contribution followed by this time-scale separation argument; it is common
practice to represent it as a singular $t=t'$ contribution leading to the
form of Eq.~\eqref{eq:ice_general}.

A common setup is shear flow, where the flow gradient can be considered
to be only along one axis, so that $\dot{\boldsymbol\gamma}$ has only one
non-trivial element, $[\dot{\boldsymbol\gamma}(t)]_{ij}=\delta_{i,x}\delta_{j,y}
\dot\gamma(t)$, where the scalar function $\dot\gamma(t)$ is called the
shear-rate of the flow. In this case, if one restrcits to
calculating just the corresponding shear-stress entry of the stress tensor,
one can reduce Eq.~\eqref{eq:ice_general} to a scalar equation,
\begin{equation}\label{eq:ice}
  \sigma(t)=\eta\dot\gamma(t) + \int_{t_0}^t G(t,t',[\dot\gamma])
  \dot\gamma(t')\,dt'\,.
\end{equation}
The scalar function $G$ is then called the generalized non-linear
shear modulus.
We work with this scalar equation in the following for simplicity;
we expect the generalization to tensorial quantities to be straight-forward.

Equations of this form in particular arise in a microscopic approach
from statistical physics, the \gls{itt} formalism \cite{Fuchs.2002c,Fuchs.2009}. In general,
the principles of perturbation theory in physics suggest that the
stress-response $\sigma(t)$ to a shear perturbation $\dot\gamma(t)$ is given
by a generalized convolution integral such as the one appearing in
Eq.~\eqref{eq:ice}. The generalized shear modulus $G$ models the material's
response.

Most common rheological models for viscoelasticity can be seen as
specializations of Eq.~\eqref{eq:ice} with specific kernels; often such
that the resulting integral equation can be converted into a
differential equation for the stress or the strain.
For example, the Maxwell model \cite{Maxwell.1867} is obtained by letting
$G(t,t',[\dot\gamma])=G_\infty\exp[-(t-t')/\tau]$ (and $\eta=0$).
We are however
interested in models where the response is more complex, and the larger
class of integral equations characterizes the response. This includes
a number of ``generalized'' Maxwell-type models, but also models derived
from the \gls{mct}, where Eq.~\eqref{eq:ice} is supplemented by another
integral equation for the evolution of $G$ \cite{FuchsBraderPRL,Brader2009}.

The form of Eq.~\eqref{eq:ice} as the result of microscopic modeling is
dictated by the fact that a shear-flow perturbation is easily modeled
as a driving term in the microscopic equations of motion -- this is
particularly true for colloidal suspensions \cite{FuchsBraderPRL,Brader2009},
but also for molecular fluids (save some thermostatting issues)
\cite{Chong.2010,Suzuki.2013}.
One thus imagines external forces to act such that a
given shear-rate $\dot\gamma(t)$ is prescribed, and the material is
characterized by the resulting shear stress $\sigma(t)$.
This decouples the Navier-Stokes equation from Eq.~\eqref{eq:ice},
and the latter is solved by numerical quadrature.
For example, linear viscoelasticity \cite{Siebenbuerger2009},
constant shear flows \cite{Fuchs2005a}, or startup-flow experiments involving
so-called stress overshoots \cite{Zausch2008,Frahsa2013} have been studied
using various \gls{mct}-derived models, or simple generalizations of
the Maxwell model \cite{Fuchs.2003}.

We are concerned with numerical solutions of the inverse problem
defined by Eq.~\eqref{eq:ice}: given a certain material stress $\sigma(t)$,
what is the strain rate $\dot\gamma(t)$ that corresponds to that imposed
stress?
An especially important example of such stress-controlled rheology is
creep, where a constant stress $\sigma(t)=\sigma_0\Theta(t)$ is fixed
for all $t>0$. The asymptotic behavior for $\dot\gamma(t\to\infty)$ is
then of interest, as it allows to distinguish flowing states
($\dot\gamma\to\dot\gamma_0>0$) from non-flowing ones ($\dot\gamma\to0$),
where the latter can be further classified into solid-like states
with finite deformation ($\gamma=\int\dot\gamma(t)\,dt<\infty$) from those
showing creeping flow ($\gamma\to\infty$ even though $\dot\gamma\to0$).
This asymptotic behavior is an important predictor of material failure.
Also the more general setting of imposed oscillatory stress,
$\sigma(t)=\sigma_0\sin(\omega t)$, is a commonly used and extremely
powerful protocol to understand materials.

The solution of this inverse problem is numerically difficult.
This is especially true if the integral kernel $G$ decays very slowly,
so that $\dot\gamma(t)$ might change non-trivially over many decades
in time, requiring adaptive-time-grid solvers.
In principle, one would also need to take into account the Navier-Stokes
equation in this case, 
but we will concentrate
on slowly varying flows where the inertial term 
can be neglected. Recall that for a flow of average density $\varrho$
and of lateral extension $L$,
one expects these inertial terms to decay on the timescale of hydrodynamic
momentum diffusion, $\tau_u=\varrho L^2/\eta$.


We discuss in this paper a robust and versatile method for the numerical
inversion of \glspl{ice}, using creep and a specific toy model of
glass-forming colloidal suspensions as an illustrative example.
The paper is organized as follows:
We first (Sec.~\ref{ch:examples}) provide a set of relevant examples
of \gls{ice} that necessitate our approach.
In Sec.~\ref{ch:classification-first_observations} we summarize some
analytical background on the type of equations given by Eq.~\eqref{eq:ice_general}.
In Sec.~\ref{ch:numerical_methods} we demonstrate the use of collocation methods as a powerful numerical technique to solve the problem. Some numerical experiments to demonstrate convergence orders and the computational effort, together with a brief application example constitute Sec.~\ref{sec:numerical_experiments}. We conclude the paper with the discussion in Sec.~\ref{sec:discussion}. More rigorous proofs for existence and uniqueness of the solution, and convergence of the collocation method in some generic rheological cases are collected in the appendix.

\section{Integral Models in Glassy Rheology}\label{ch:examples}

Let us give some examples with physical applications that fit
the class of models we consider.
\begin{example}[Kelvin-Voigt model]
This is the trivial case where
\begin{equation}
G_\text{KV}(t, t', [\dot{\gamma}]) = G_\infty
\end{equation}
is constant. In the literature this model is usually given in the form of an
equivalent differential equation for
$\gamma(t) = \int_{t_0}^t \dot{\gamma}(s) ds$:
\begin{equation}
\sigma(t) = G_\infty \gamma(t) + \eta \dot{\gamma}(t) ; \quad \gamma(t_0) = 0\,.
\end{equation}
The model describes a visco-elastic solid with shear modulus $G_\infty$.
Its creep response is an exponentially relaxing shear rate with time
scale $\tau_V=\eta/G_\infty$,
\begin{equation}\label{eq:KVcreep}
  \dot\gamma_\text{creep,KV}(t)=\frac{\sigma_0}{\eta}e^{-t/\tau_V}\,.
\end{equation}
Since the integrated strain $\gamma<\infty$ in this case, this is the
response of a solid.
\end{example}
\begin{example}[Jeffreys model]
Another central model in the literature is the Jeffreys model for
a visco-elastic fluid, given by
\begin{equation}
G_\text{J}(t, s, [\dot{\gamma}]) = G_\infty e^{-\frac{t-s}{\tau}}
\end{equation}
Also this model can be formulated in terms of a a differential equation,
\begin{align}\label{eq:Jode}
\dot{\sigma}(t) + \frac{1}{\tau} \sigma(t) = \bigg( G_\infty + \frac{\eta}{\tau} \bigg)  \dot{\gamma}(t) + \eta \ddot{\gamma}(t); \quad \dot{\gamma}(0) = \frac{\sigma(0)}{\eta}\,.
\end{align}
In the case $\eta = 0$ this is Maxwell's model.
The creep response of the Jeffreys model is
\begin{equation}\label{eq:Jcreep}
  \dot\gamma_\text{creep,J}(t)=\frac{\sigma_0}{\eta_\text{M}}
  +\frac{\sigma_0}{\bar\eta}e^{-t/\tau_\text{J}}
\end{equation}
with $\tau_\text{J}=\eta\tau/\eta_\text{M}$, $\eta_M=G_\infty\tau+\eta$,
and $\bar\eta=\eta(1+\eta/G_\infty\tau)$. In the case $\eta\to0$,
the last term becomes a singular contribution around $t=0$, and Maxwell's
result $\dot\gamma=\sigma_0/G_\infty\tau$ is recovered.
Since $\dot\gamma(t)$ approaches a constant, this is the response of a
viscous fluid.
\end{example}

Many complex materials do not show the clear single-exponential relaxation
that is predicted by the Jeffreys or Kelvin-Voigt models, but instead
show broad relaxation-rate spectra that are well described by power laws.
This motivates the introduction of fractional viscoelastic models, often
formulated in terms of the Caputo fractional derivative ($n\in\mathbb N_0$
and $0<\alpha<1$),
\begin{equation}
  \frac{\partial^{n+\alpha}}{\partial t^{n+\alpha}}f(t):=\frac1{\Gamma(1-\alpha)}
  \int_{t_0}^t\frac{f^{(n+1)}(t')\,dt'}{(t-t')^\alpha}\,.
\end{equation}

\begin{example}[Fractional Viscoelasticity]\label{ex:fractional-viscoelasticity}
The most basic model of fractional viscoelasticity interpolates between fluid-
and solid-like response by letting \cite{Bonfanti2020}
\begin{equation}
  \sigma(t) = \tilde G \frac{\partial^\beta}{\partial (t/\tau)^\beta}\gamma(t)
\end{equation}
with some internal relaxation time $\tau$,
such that $\beta=0$ corresponds to the Hookean solid, and
$\beta=1$ to the Newtonian fluid.
This is, in fact, an integral model, setting
\begin{equation}
  G_\text{FVE}(t,s,[\dot\gamma])=\tilde G\frac{((t-s)/\tau)^{-\beta}}
  {\Gamma(1-\beta)}
\end{equation}
(and $\eta=0$ although a background viscosity can be easily included).
The creep response of this model (for $\eta=0$) is easily calculated
from the Laplace transform; one obtains (for $0<\beta<1$)
\begin{equation}
  \dot\gamma_\text{creep,FVE}(t)=\frac{\sigma_0}{\Gamma(\beta)}
  \frac1{\tilde G\tau^\beta}t^{\beta-1}\,.
\end{equation}
\end{example}

The models discussed so far are all linear rheological models, in the
sense that they do not incorporate a nonlinear dependence of the
stress response to a shear-rate perturbation.
The description of
materials with non-linear response, in particular yield-stress materials,
motivates modifications of the simple viscoelastic models. In particular,
the modification of Eq.~\eqref{eq:Jode} with a strain-rate-dependent
(or stress-dependent) relaxation rate $1/\tau$ defines a class of
models, called White-Metzner models.

\begin{example}[Accumulated strain model]
This White-Metzner-type model, originally suggested by
Bogue \cite{Bogue1966}, lets the stress relaxation depend on the
accumulated strain $\bar\gamma_{tt'}=\int_{t'}^t |\dot{\gamma}(s)| ds$,
\begin{equation}
G_\text{AS}(t, t', [\dot{\gamma}]) = G_\infty e^{-\frac{t-t'}{\tau}} e^{-\int_{t'}^t \frac{|\dot{\gamma}(s)|}{\gamma_c} ds }
\end{equation}
with a material parameter $\tau \in \mathbb{R}_{+} \cup {\infty}$ that sets the history-independent stress relaxation, and a strain-scale parameter
$\gamma_c>0$.
The model describes shear thinning, since the effective relaxation rate
of stress fluctuations (quantified through $G(t,t')$), increases with
increasing shear rate. It was rediscovered in this context as a sketch of
\gls{mct} and termed ``generalized nonlinear Maxwell model'' \cite{Fuchs.2003}.
The corresponding \gls{ode} reads
\begin{equation}
\dot{\sigma}(t) + \bigg(\frac{1}{\tau} + \frac{|\dot{\gamma}(t)|}{\gamma_c} \bigg) \sigma(t) = \bigg( G_\infty + \frac{\eta}{\tau} + \eta \frac{|\dot{\gamma}(t)|}{\gamma_c} \bigg)  \dot{\gamma}(t) + \eta \ddot{\gamma}(t);
\end{equation}
with $\dot{\gamma}(0) = \frac{\sigma(0)}{\eta}$.
Note, in the limits $\gamma_c \rightarrow \infty$ and $\gamma_c \rightarrow 0$ one recovers the Maxwell model and the Newtonian fluid respectively.
\end{example}

This example of a nonlinear model in fact reduces the \gls{ice} \eqref{eq:ice}
to an \gls{ode}.
However, this might not be sufficiently generic to capture complex-flow
phenomena in glass-forming fluids. For example,
\citeauthor{Siebenbuerger2012} \cite{Siebenbuerger2012}
studied the emergence of logarithmic creep in colloidal suspensions
and suggested a model where the stress relaxation is driven by the
instantaneous shear rate rather than the accumulated strain.

\begin{example}[Strainrate Relaxation model]\label{ex:SRR_introduction}
Following Ref.~\cite{Siebenbuerger2012} we consider
\begin{equation}\label{eq:srrG}
G_\text{SRR}(t, t', [\dot{\gamma}]) = G_\infty e^{-\frac{t-t'}{\tau}} e^{-\frac{(t - t')|\dot{\gamma}(t')|}{\gamma_c} }\,.
\end{equation}
This model has a non-analytic dependence on the strain rate at the
point $\dot\gamma=0$.
\end{example}
A remark on the appearance of $|\dot\gamma|$ in Eq.~\eqref{eq:srrG}
might be in order. One expects the full strain-rate tensor to enter
any rheological model through its tensorial invariants, in order to
ensure that the resulting model is compatible with the frame-invariance
principles of classical continuum mechanics. The absolute value
$|\dot\gamma(t)|$ arises in fact from the second invariant of the tensor
$\dot{\boldsymbol\gamma}(t)$ under stationary shear.

The examples given above are in particular relevant for the rheology
of yield-stress fluids: in the case of stationary shear, $\dot\gamma(t)
=\dot\gamma=\text{const.}$ (with $t_0\to-\infty$), the resulting
stationary stress $\sigma(\dot\gamma)$ approaches a finite constant,
the dynamical yield stress, $\sigma(\dot\gamma)\to\sigma_y>0$, as
$\dot\gamma\to0$. This immediately arises from Eq.~\eqref{eq:ice}
if $G(t,t';\dot\gamma)$ becomes a function of $\dot\gamma(t-t')$
under these circumstances. Both the accumulated strain, and the
strainrate-relaxation model fulfill this criterion. In fact, they
are equivalent under the imposition of constant strain rate, and offer
different physically motivated generalizations to time-dependent shear.

Another hallmark of glass-forming fluids is the appearance of
stress overshoots in the startup flow: after imposing a constant shear
rate, $\dot\gamma(t)=\dot\gamma_0\Theta(t)$, one observes the stress
$\sigma(t)$ to increase non-monotonically to its stationary value.
Crucially, such stress overshoots can be observed in homogeneous-flow
conditions, without the necessity of shear bands \cite{Zausch2008}.
In terms of the \gls{ice} framework, they thus require to model
a dynamical shear modulus that becomes transiently negative as a function
of the imposed strain. Theories like \gls{mct} suggest to separate
this as an effect of a nonlinear response of the elastic coefficients,
from the plastic relaxation contributions; effectively one keeps the
exponential relaxation for the latter and imposes an empirical
function for the strain-dependent $G_\infty(\gamma)$.
\begin{example}[Stressovershoot model]\label{ex:SO-introduction}
In the empirical model
proposed by \Citeauthor{Amann2013} \cite{Amann2013} and
\Citeauthor{Frahsa2013} \cite{Frahsa2013}
one sets
\begin{equation}\label{eq:soG}
  G_\text{SO}(t,t',[\dot\gamma]) = \frac{G_\infty\big(\gamma_{tt'}[\dot{\gamma}]\big)}{G_\infty}
    G_\text{SRR}(t,t',[\dot\gamma])
\end{equation}
where $\gamma_{tt'}[\dot{\gamma}]=\int_{t'}^t\dot\gamma(s)\,ds$ is the total
strain%
\footnote{Note that this differs from the accumulated strain $\bar\gamma_{tt'}$
in cases where $\dot\gamma(t)$ changes sign. To describe elastic deformation, the total strain is appropriate, while for accumulated strain-induced plastic deformations, $\bar\gamma$ is used in the relaxation function.}
and the strain-softening function
\begin{equation}
  G_\infty(\gamma)=G_\infty
  \left(1-\left(\frac{\gamma}{\gamma_\ast}\right)^{2n}\right)
  e^{-\left(\frac{\gamma}{\gamma_{\ast\ast}}\right)^{2n}}\,.
\end{equation}
Here $\gamma_{\ast}$, $\gamma_{\ast\ast}$, and $n\in\mathbb N$ are parameters
that are adjusted empirically to match position, strength and shape of
the stress overshoot in experimental data \cite{Amann2013}.
\end{example}
We will be using the \gls{so} model to elucidate the numerical scheme
that we develop in the following.

For completeness, let us briefly mention another widely used class
of nonlinear rheological models, the K-BKZ models. They are commonly
formulated in terms of the strain, and not the strain rate, as a fundamental
model function. However, they fall into the same category as the models
we discuss.
\begin{example}[Scalar K-BKZ model]\label{ex:KBKZ}
The standard K-BKZ model \cite{Kaye1962} in simple shear 
is given by
\begin{equation}
\sigma(t) = \int_{t_0}^t m\big(t - s, \gamma(s)\big) \, \gamma(s) ds \,.
\end{equation}
Identifying $G(t,s,[\dot\gamma])=\int_s^t m(t-s',\gamma(s'))\,ds'$ allows
to cast the model into the form of Eq.~\eqref{eq:ice}, assuming that
integration by parts is allowed.
(In this definition, $G(t,t)=0$, which is different than the
Maxwell-type shear modulus that would imply $G(t,t)=G_\infty$, since
an instantaneous strain contribution is treated differently in the K-BKZ
model.)
The assumption of a proper convolution in the K-BKZ models,
$m(t-s,\gamma(s))$ instead of $m(t,s,\gamma(s))$, rules out some
nonlinear rheological features in comparison to the models that we
discussed above \cite{Voigtmann.2012}.
\end{example}

\section{Classification}\label{ch:classification-first_observations}

As shown by the examples above, a number of relevant rheological models
cannot be formulated as \glspl{ode}. Thus, the fundamental
evolution equation Eq.~\eqref{eq:ice} remains an integral equation. 
Specifically, if the dependence of the generalized shear modulus on the
shear rate is time-local, $G(t,t',[\dot\gamma])=G(t,t',\dot\gamma(t'))$,
one obtains a so-called Volterra integral equation.
The theory of Volterra equations is rich; we refer the reader to the
books by \citeauthor{Hackbusch1995} \cite{Hackbusch1995} and
by \citeauthor{Wazwaz2011} \cite{Wazwaz2011} as introductory texts,
and to \citeauthor{Gripenberg1990} \cite{Gripenberg1990} for a more
thorough and abstract treatment.
One distinguishes three types of Volterra integral equations:
denoting the unknown function to be determined as the solution of the
problem by $u(t)$, the general Volterra equation
reads
\begin{equation}
a(t) u(t) = f(t) + \int_{t_0}^t k(t, s, u(s))\,ds
\end{equation}
In the case where $a(t)\equiv0$, one speaks of a Volterra equation
of the first kind. Setting $a(t)=\text{const.}$ (and thus $a=1$ without
loss of generality) one obtains a Volterra equation of second kind,
while the fully general case is referred to as a Volterra equation
of the third kind.
Here $k(t, s, y)$ is the kernel of the equation and $f(t)$ is sometimes refered to as the forcing function. In the Volterra equation of third kind the factor $a(t)$ needs to be a continuous function with only a finite number of zeros in the interval $[t_0, T]$ over which the solution is sought. 
We speak of linear (nonlinear) Volterra equations if the kernel $k$
is linear (nonlinear) in $u(s)$.
Some nonlinear equations that have been studied include the
specific nonlinear version (see, e.g., Ref.~\cite{Zhang.2016})
\begin{equation}
  u(t)=f(t)+\int_{t_0}^tk(t,s,u(t),u(s))\,ds
\end{equation}
or the equations of \emph{autocorrelation type},
\begin{equation}\label{eq:avie}
  u(t)=f(t)+\int_{t_0}^tk(t,s)u(t-s)u(s)\,ds\,,
\end{equation}
termed \gls{avie}.
Such autoconvolution equations are known to be ill-posed, in the sense
that the solution might change drastically with small changes in the
parameters (i.e., they are prone to show bifurcations).

To connect the rheological models to mathematical theory, we identify
the strainrate as $\dot\gamma(t)=u(t)$, and the stress as the forcing
function, $\sigma(t)=f(t)$. Thus,
\begin{equation}
\eta u(t) = \sigma(t) - \int_{t_0}^t G\left(t, s, [u] \right) u(s) ds := H[u](t)\,.
\label{equ:abstract-Volterra}
\end{equation}
Adopting the classification of Volterra equations, we distinguish
models without background viscosity ($\eta=0$) as \gls{ice} of first kind,
and those with a Newtonian contribution ($\eta\neq0$) as \gls{ice}
of second kind.
In rheology terms, this amounts to the distinction between the Maxwell and the
Jeffreys model.
An \gls{ice} of the first kind can in principle be transformed into one
of second kind by taking a derivative,
re-identifying the forcing function with $\dot\sigma(t)$; we do not discuss
this case separately.

In the general rheological \gls{ice}, the dependence of the shear modulus
on the strain rate is not time-local; in this general case of a
functional dependence, one speaks of $H[u]$ in Eq.~\eqref{equ:abstract-Volterra}
as an abstract Volterra operator.
Causality implies that it depends only on the past values $u(s)$ with
$s\in[t_0,t]$;
such operators $H$ are also called \emph{causal} or \emph{anticipative}
(cf.~\cite[Ch.12 sec. 2 and 8]{Gripenberg1990}).

Let us elucidate the classification of the relevant examples given above:
The \gls{fve} model, Example~\ref{ex:fractional-viscoelasticity},
defines a linear Volterra equation (of first kind in the absence of a
Newtonian background viscosity). Its integral kernel is weakly singular;
such equations are also called of integral equations of Abel type.
The K-BKZ model, Example~\ref{ex:KBKZ}, presents a Volterra
equation, either by the transformation discussed there,
or directly by identifying the strain as the function to
solve for, $u(t)=\gamma(t)$. These models are thus subject to the same
mathematical analysis and the same numerical treatment.

The \gls{srr} model, Example~\ref{ex:SRR_introduction}, with (without)
a Newtonian background viscosity defines a nonlinear Volterra equation of
second (first) kind.
The \gls{so} model, Example~\ref{ex:SO-introduction}, does not define
a Volterra integral equation due to the dependence of the shear modulus
on the accumulated strain, and not just the instantaneous strain rate.
But it does define a causal abstract Volterra operator, as introduced above.

In particular, we can leverage the theory of Volterra integral equations
to assert existence and uniqueness of solutions of the rheological models.
For the models that lead to standard Volterra equations, such as the
\gls{srr} model, this follows directly from the standard theorems
for Volterra equations of the second kind with regular kernels. We briefly
summarize the relevant results and apply them to the \gls{srr} model
as an example in Appendix~\ref{app:nonlinear-Volterra}.
For the \gls{fve} model, whose kernel is weakly singular,
an existence and uniqueness theorem can be found, for example,
in the book by \citeauthor{Brunner1986} \cite[comments below Theorems 1.3.8 and 1.3.9]{Brunner1986}. There, also existence and uniqueness theorems
for Volterra equations of the first kind with regular and weakly singular
kernels can be found \cite[Theorems 1.3.11 and 1.3.12]{Brunner1986}.

As noted, the \gls{so} model is more general; here one can however rely
on the fixpoint theorems for general functional equations, see for
example \citeauthor{Gripenberg1990} \cite[Ch.~12, Theorem 12.1,
Corollaries 12.2 and 12.3]{Gripenberg1990}.
Since the model (and its variants) have some relevance in the
study of glass-forming fluids, we establish existence and uniqueness
of (continuous) solutions for this model in Appendix~\ref{app:functional-equations}.

\section{Numerical methods}\label{ch:numerical_methods}

We examine two numerical approaches to equations of the type (\ref{equ:abstract-Volterra}), using as examples the \gls{srr} model (see example \ref{ex:SRR_introduction}) and the \gls{so} model (see example \ref{ex:SO-introduction}).
In particular we extend standard methods that were originally
developed for Volterra equations to the nonlinear Volterra operators
for which the \gls{so} model is a physically motivated example.
A particular emphasis is on collocation methods, introduced below
in Sec.~\ref{sec:Collocation}. The models that we study have a unique
solution, as ascertained by the standard techniques for Volterra
equations (see Appendix~\ref{app:nonlinear-Volterra}), extended to
the functional equations of type \eqref{equ:abstract-Volterra} in
Appendix~\ref{app:functional-equations}.
For the collocation methods, we are able to
prove convergence of the numerical scheme
(see Appendix~\ref{app:convergence-proof}), and this
serves as a reference point also for simpler, more straight-forward
quadrature schemes. To our knowledge, such a proof
has not appeared yet in the mathematical literature.

\subsection{Preliminaries}

The theory for the numerical solution of Volterra-integral equations is well developed and we follow these standard techniques.
We refer to the introductory texts by \citeauthor{Hackbusch1995} \cite{Hackbusch1995} or by \citeauthor{Linz1985} \cite{Linz1985}; a more extensive
treatment is given by \citeauthor{Brunner1986} \cite{Brunner1986}.

Our goal is to numerically approximate the solution of Eq.~\eqref{equ:abstract-Volterra} of second type (with $\eta \neq 0$)
on an Intervall $[t_0, T]$. To do that we replace the interval
by a grid $(t_i)_{i = 0, \dots, N}$ of points, 
\begin{align}
&t_0 < t_1 < \dots < t_N \leq T \,,
\\ & h_i := t_{i+1} - t_i \,.
\end{align}
Now, we need to determine the approximations $u_i \approx u(t_i)$ to the solution on the grid points $t_i$.  (For any function, we will denote $f(t_i)=f_i$ for brevity if convenient.)
The most direct ansatz for a numerical method is to approximate the time-integrals $\int_{t_0}^{t_i} G(t_i, s, [u]) u(s)\, ds$ by some quadrature rule using the grid points as nodes. This approach will be shown in section \ref{sec:DQ}. Another approach that has proven fruitful in many areas of numerics, is the collocation ansatz, where the solution is approximated by piecewise polynomial functions. This approach will be followed in section \ref{sec:Collocation}. 
Both approaches can be refined by applying Richardson extrapolation, as found in introductory numerics books (e.g. \cite[Ch.3.5]{Stoer2002})). For convenience we will outline the main ideas in 
subsection \ref{sec:DQ} with view towards an application together with direct quadrature methods. 
The approaches we show here are all formulated for \textit{regular} kernels $G(t, s, [u])$. Fractional models of the form $G(t, s, [u])= \tau^\beta (t-s)^{-\beta} \tilde{G}(t, s, [u])$ require some extra care, although
the ideas of using a quadrature or collocation ansatz survive at least in the case of Volterra equations and are variously called {fractional convolutional quadrature}, {fractional VLM methods}, {collocation on graded meshes}, or {collocation with nonpolynomial spline spaces}; see for example \cite[Ch. 6]{Brunner1986} for more details. 

The numerical methods presented here require the exact solution $u$ to be at least continously differentiable ($ u \in C^1([0, \infty))$). For this it is necessary, that the forcing function $\sigma(t)$ is also at least $C^1$. Especially jumps in $\sigma(t)$, $t \in [0, \infty)$, are strictly speaking not allowed and need to be treated separately. 

\subsection{Direct Quadrature ansatz}\label{sec:DQ}

\begin{align}
\eta u_n &= \sigma_n - \int_{t_0}^t G(t, s, [u]) u(s) \, ds 
\\ &\approx \sigma_n - \sum_{i = 0}^{n} w_{n, i} G(t_n, t_i, [u]) u_i
\end{align}
where $w_{n,i}$ are the weights of the quadrature rule.
We keep the notation $[u]$ to denote a potentially functional dependence
on $u$. Since we assume $G$ to be causal, for any given quadrature point
this implies a potential dependence on all $(u_j)_{j=0,\ldots n}$.
The three most elementary examples are
\begin{example}[Quadrature rules]
Setting $\Delta t_i=t_{+1}-t_i$,
the well known left-point (LP), right-point (RP), and trapezoidal (TZ) quadrature rules,
\begin{subequations}
\begin{align}
\int_{t_{i}}^{t_{i+1}} G(t_n, s, [u]) \, u(s)\,ds &\approx \Delta t_i G(t_n, t_i, [u])\, u_i \\
\int_{t_{i}}^{t_{i+1}} G(t_n, s, [u]) \, u(s)\,ds &\approx \Delta t_i G(t_n, t_{i+1}, [u])\, u_{i+1} \\
\int_{t_i}^{t_{i+1}} G(t_n, s, [u]) \,u(s)\,ds &\approx \frac{\Delta t_i}{2}\left(  G(t_n, t_{i+1}, [u])\,u_{i+1} \right.\\ \nonumber & \qquad\left.- G(t_n, t_i, [u])\,u_i \right)
\end{align}
\end{subequations}
yield, respectively,
\begin{subequations}
\begin{align}
\eta u_n &= \sigma_n - \sum_{i = 0}^{n-1} h_i G(t_n, t_{i}, [u])\, u_{i}\\
\eta u_n &= \sigma_n - \sum_{i = 0}^{n-1} h_i G(t_n, t_{i+1}, [u])\, u_{i+1}\\
\eta u_n &= \sigma_n - \sum_{i = 0}^{n-1} \frac{h_i}{2}\left(  G(t_n, t_{i+1}, [u])\,u_{i+1}  + G(t_n, t_i, [u])\,u_i \right)
\end{align}
\end{subequations}
\end{example}
In the case of a Volterra equation of second kind, $G(t,s,[u])=G(t,s,u(s))$,
the choice of quadrature rule is sufficient to define the numerical scheme.
Higer-order quadrature rules could in principle be used, but they necessarily
lead to multistep methods that cannot start on their own, and they become
more intricate on nonlinear grids $(t_i)_{i=0,\ldots N}$. For this reason
we do not consider higher-order direct quadrature methods here.

Note that for Volterra equations, there is also a close correspondence of
these quadrature methods to linear multi-step methods for \glspl{ode}:
the leftpoint method corresponds to an explicit Euler method, the rightpoint
method to an implicit Euler method, and the trapezoidal rule corresponds
to the trapezoidal method for \glspl{ode}.%
\footnote{To see this consider $G(t, s, u(s)) = G(s, u(s))$ to be independent on $t$. Then the integral equation is equivalent to a differential equation and one can show that the algorithms yield the explicit or implicit Euler methods respectively.}.
Such quadrature rules are also
called $(\rho,\sigma)$-reducible. At least for linear Volterra equations,
$G(t,s,u(s))u(s)=G(t,s)u(s)$, one can show that if the underlying \gls{ode}
method has certain properties, the resulting algorithm applied to the
integral equation of first kind ($\eta=0$) converges with the same order
\cite{Wolkenfelt1981}. The three methods outlined above fulfill the
required properties. The order of convergence $p$ is given by the leading
order of the error in terms of the step size, $h^p$.
The leftpoint rule for a Volterra equation of second kind leads to
a method of order $p=1$, while the trapezoidal rule results in a method
of order $p=2$ \cite[Corrolary 3.2.1]{Brunner1986}.

For a Volterra equation it is sufficient to choose a quadrature rule.
However, in the case of more general operators $G(t, s, [u])$ one needs to
additionally chose how to approximate $G(t_n, t_i, [u])$ at the corresponding
grid points, through representing the functional dependence on the numerical
grid. This is a potentially crucial detail: $G(t_n,t_i,[u])$ might involve
$u_n$ so that the chosen method necessarily becomes an implicit method.
Nonlinear integral equations with such functional dependence can show
bifurcations in the $t=\infty$ limit so that a proper treatment of the
implicit term is crucial to obtain solutions that are asymptotically
correct as $t$ becomes arbitrarily large
-- in a related context, \gls{mct} gives
a mathematically intricate example \cite{Goetze.1995b}.

\begin{figure}
	\centering
	\includegraphics[width = 0.8\linewidth]{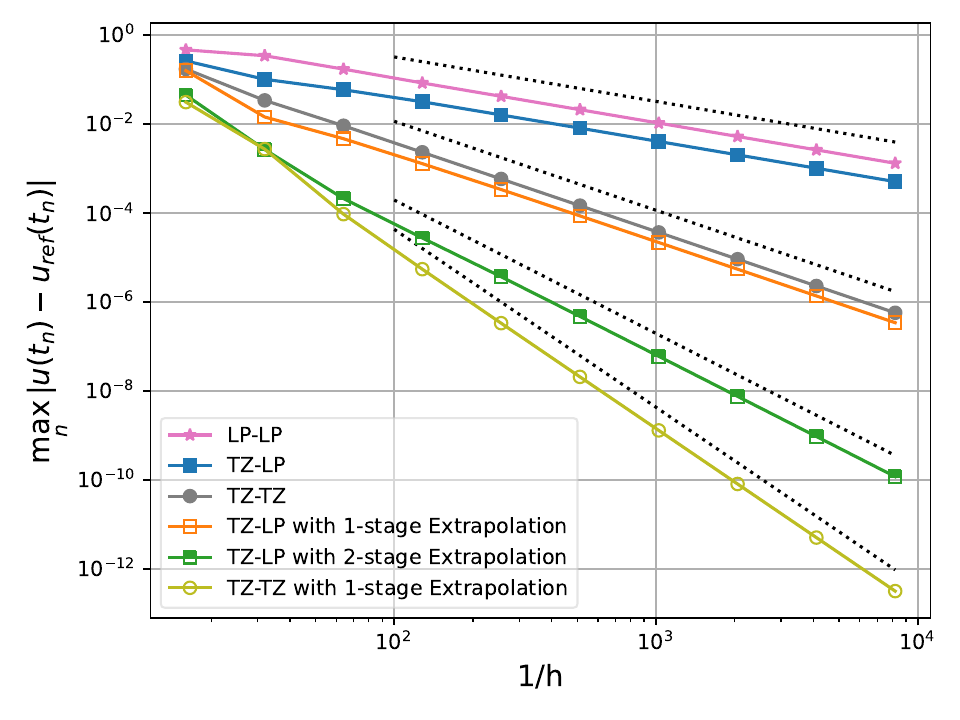}
	\caption{Convergence test for the direct quadrature methods, applied to the test case of the stressovershoot model described in Sec.~\ref{sec:testcase}.
	The error norm relative to the reference exact solution
	(see Sec.~\ref{sec:testcase}) is shown as a function of the inverse steplength $1/h$. The dotted lines represent the slopes $h^p$, $p = 1, 2, 3, 4$ respectively.
	The methods are labeled by their quadrature rule for the
	outer integral followed by that used for the inner integral,
	cf.\ Eq.~\eqref{eq:SO-TZ-LP} as an example.
	}
	\label{fig:DQ_u1-SO-a}
\end{figure}

The direct quadrature methods for models like the \gls{so} model
are characterized by their
``outer'' quadrature scheme (used to approximate
$\int_{t_0}^tG(t,s,[u])\,ds$) supplemented by the ``inner'' scheme
(used to approximate $z(t,s)=\int_s^t u(t')\,dt'$. These
need not be the same, and we name them outer--inner accordingly
(for example, leftpoint--leftpoint (LP-LP) etc.)
\begin{example}[Stressovershoot model]
In the \gls{so} model, Example~\ref{ex:SO-introduction}, $G(t,s,[u])$
for any interval $[s,t]$ 
is of the form
\begin{equation}
G(t, s, [u]) = G(t, s, {\textstyle\int_{s}^t u(t') dt'}, u(s))
\end{equation}
We can choose a direct quadrature approximation for the strain integral
$z(t,s)$,
\begin{equation}
z_{n, i} = \sum_{l = i}^n w_{n,l} u_l \approx z(t_n, t_i) = \int_{t_i}^{t_n} u(t') dt'\,.
\end{equation}
and the trapezoidal rule for the outer integral.
Thus, the trapezoidal--leftpoint scheme is obtained,
\begin{subequations}\label{eq:SO-TZ-LP}
\begin{align}
\eta u_n &= \sigma_n - \sum_{i = 0}^{n-1} \frac{h_i}{2}\left(  G(t_n, t_{i+1}, z_{n,i+1}, u_{i+1})\,u_{i+1}\right. \\ \nonumber &\left.  + G(t_n, t_i, z_{n,i} ,u_i)\,u_i \right)\,, \\
z_{n, i} &=\sum_{j = i}^{n-1} h_j u_j\,,
\end{align}
\end{subequations}
and similar schemes follow by analogy.
The leftpoint--leftpoint method for $u_n$ is fully explicit,
while the trapezoidal--leftpoint method is implicit but leads to a
linear system for $u_n$. The trapezoidal--trapezoidal method is
fully implicit and $u_n$ is determined by the solution of a (typically)
nonlinear equation.
Our experimental convergenve analysis (see Fig.~\ref{fig:DQ_u1-SO-a}) indicates that the former two methods are first order methods, while the third method appears to be of second order as expected from the orders of the underlying quadrature rules\footnote{The numerical testcase (sec. \ref{sec:testcase}) was chosen such that the nonanalyticity given by the $\exp(-|u(s)|(t-s)/\gamma_c)$-term did not become relevant. However, even if the exact solution $u(s)$ changes sign, we would still observe a 'rugged' second order convergence for the TZ-TZ method, but one that cannot be improved by Richardson extrapolation (see Appendix \ref{appendix:trapezoidal_jumps}).}.
\end{example}

The convergence order of the given quadrature rule can be improved
by a standard method called Richardson extrapolation
(see, e.g., Ref.~\cite[Ch.~3.5]{Stoer2002}).
The idea rests on the observation that in many cases the error
of the numerical solution $\tilde u_h(t)$ with respect to the true
solution $u(t)$ can be expanded as a polynomial in the step size $h$
as $h\to0$:
\begin{equation}
u(t) - \tilde{u}_h(t) = c_1(t) h^p + c_2(t) h^{p+1} + \ldots 
\end{equation} 
where $p$ is the order of convergence. Interpreting $\tilde u_h(t)$ as
a polynomial $P_R(h)$ in $h$, the exact solution is $P_R(0)$ and can be
approximated from knowledge of the values $P_R(h)$ and $P_R(h/2)$. This
defines the improved numerical solution
\begin{align}
u_R(t) = P_R(0) &= \alpha \tilde{u}_h(t) + (1 - \alpha) \tilde{u}_{h/2}(t)\,,
 & \alpha &= \frac{1}{1 - 2^p}
\end{align}
as an approximation of order $O(h^{p+1})$ to the exact solution.
This scheme can be iterated if more higher-order error terms are known.
For the trapezoidal rule, the error can be developed in terms of
$h^{2n}$ ($n\in\mathbb N$), however this is only true for the quadrature
method if the integrand is smooth enough (see Appendix~\ref{appendix:trapezoidal_jumps}).
In the physical problem at hand, this can be guaranteed for problems where
the shear rate never changes sign, due to the appearance of $|\dot\gamma(t)|$
in some of the models.

Richardson extrapolation can be generalized to non-uniform grids: it is
sufficient that the expression for the error in terms of the step length
factorizes the time- and stepsize-dependence, so that the Richardson
scheme can be applied by using the solution on a given grid, and the
grid that is defined by subdividing each grid intervall by two.

Figure~\ref{fig:DQ_u1-SO-a} shows the performance of several direct quadrature
methods for the \gls{so} model, i.e., a model of generalized Volterra-operator
type. Numerical details for the model and the used grids are given below
in Sec.~\ref{sec:testcase}.
One observes the expected orders of convergence, $p=1$ for the LP rule,
$p=2$ for the TZ rule, with the quadrature of the lower order dictating
the convergence for the combined scheme.
Richardson extrapolation increases the order of convergence by one for
methods involving the LP rule, and by two if the extrapolation is iterated;
for the TZ rule, each extrapolation gains two orders of convergence.
Thus the TZ-TZ scheme with Richardson extrapolation is a fourth-order
scheme.

\subsection{Collocation ansatz}\label{sec:Collocation}

Collocation methods are close cousins of the well-known Runge-Kutta
schemes for the numerical solution of (stiff) \glspl{ode}.
An extensive introduction into collocation methods applied to Volterra equations can be found in the book by \citeauthor{Brunner2004} \cite{Brunner2004}.

In the following we will outline the principal idea and show how to apply this to derive methods for more general types of evolutionary integral equations.

The essential idea of an collocation ansatz is to approximate the exact solution $u$ by a spline function $p$, e.g. by a piecewise polynomial function 
\begin{align}
u(t) &\approx p(t) = p_i(t) = \sum_{j = 0}^{m} a_{i, j} P_j(t), \quad t \in (t_{i}, t_{i+1}]
\\ u(t_i) &\approx p_{i-1}(t_i)
\end{align}
of some fixed maximum degree $m$, where the knots correspond to the grid-points $t_i$, $i = 0, \dots, N$ and $P_j(t)$, $j=(1,\ldots m)$, is some
polynomial basis.
A usual choice are the monomials, $P_j(t)=t^j$,
but in the present case we found it more stable to choose a Lagrange-polynomial
basis $P_j(t)=L_{j+1}(t)$.

Additionally, for each interval $[t_i, t_{i+1}]$ one defines a set of \textit{collocation points} $\tilde{t}_{i, k} = t_i + c_k h_i $, $k = 1, \dots, m+1$ based on the \textit{collocation parameters}
\begin{align}
0 \leq c_1 < \dots < c_{m+1} \leq 1 .
\end{align} 
In the numerical analysis of Volterra equations it is common to use discontinous splines that may have jump discontinuities at the knots $t_i$.
Thus, on each interval $[t_i, t_{i+1}]$ the collocation polynomial is uniquely determined by its values on the $m+1$ collocation points $\tilde{t}_{i,k}$, $k = 1 \dots m+1$.  
Demanding that the collocation polynomials solve the given problem exactly
on the collocation points
defines a nonlinear system
of equations for the coefficients $a_{i,j}$ of the polynomials.
Applied to Eq.~\eqref{equ:abstract-Volterra}, the
\emph{collocation equation} reads
\begin{equation}
\sigma(\tilde{t}_{i,k}) = \eta p_i(\tilde{t}_{i,k}) + \int_{t_0}^{\tilde{t}_{i,k}} G(t, s, [p]) p_i(s) ds, \quad i \in \{0, \dots, N-1\}; k \in \{1, \dots, m+1 \}\,.
\end{equation}
Note that depending on the choice of the $p_i(t)$, it can even be possible
to evaluate the integral analytically, as demonstrated by the example below.
This offers a nice advantage for some models and can be helpful to capture
properties of the integral that might be difficult to restore purely by
numerical quadrature unless the time step is chosen very small.
Such applications of simple variants of collocation methods are already
well known from, e.g., Fouier-Laplace transforms of numerical data over
wide frequency bands \cite{Tuck.1967}, or for the evaluation of
\gls{mct} memory kernels in two spatial dimensions where singular Jacobians
appear due to the desired variable transform \cite{Caraglio.2020}.

\begin{example}[Collocation in the \gls{srr} model using piecewise constant splines]\label{ex:Collocation_SRR}
A particularly simple collocation method is obtained by letting $p_i(t)=p_i$
be constant. Then,
only one collocation parameter $c_1 = \theta \in (0, 1]$ is needed.
For the \gls{srr} model, Example~\ref{ex:SRR_introduction}, with $\tau = \infty$, the integrals can then be performed analytically and lead to the collocation equation
\begin{align}
\sigma(t_i + \theta h_i) &= \eta p_i + \int_{t_i}^{t_i + \theta h_i} G_\infty e^{- \frac{(t_i + \theta h_i - s) |p_i|}{\gamma_c} } p_i ds + \sum_{j=0}^{i-1} \int_{t_j}^{t_{j+1}} G_\infty e^{-\frac{(t_i + \theta h_i - s) |p_j| }{\gamma_c}} p_j ds \nonumber
\\&= \eta p_i + G_{\infty} \gamma_c \frac{p_i}{|p_i|} \left(1 - e^{-\frac{\theta h_i |p_i|}{\gamma_c}} \right) 
\\&\quad+ \sum_{j=0}^{i-1} G_{\infty} \gamma_c \frac{p_j}{|p_j|} \left( e^{-\frac{(t_i+\theta h_i - t_{j+1})|p_j|}{\gamma_c}} - e^{-\frac{(t_i+\theta h_i - t_{j})|p_j|}{\gamma_c}}\right)\,, \nonumber 
\end{align}
defining an implicit equation for the coefficients $p_i$. 
By Identifying 
\begin{align}
u(t_{i+1}) \approx u_{i+1} = p(t_{i+1}) = p_{i}(t_{i+1}) = p_i
\end{align}
and rearranging some terms the derived collocation method reads 
\begin{align}
u_{i+1} &= \frac{\sigma(t_i + \theta h_i)}{\eta} - \frac{G_\infty \gamma_c}{\eta}\frac{u_{i+1}}{|u_{i+1}|} \bigg( 1 - e^{- \frac{\theta h_i |u_{i+1}|}{\gamma_c} }\bigg) 
\\&\quad + \frac{G_\infty \gamma_c}{\eta} \sum_{j = 0}^{i-1} \frac{u_{j+1}}{|u_{}j+1|} \bigg( e^{-\frac{(t_i+\theta h_i - t_{j+1})|u_{j+1}|}{\gamma_c}} - e^{-\frac{(t_i+\theta h_i - t_{j})|u_{j+1}|}{\gamma_c}}\bigg) 
\end{align}
This yields a first order method for all choices of $\theta \in (0, 1]$
(while $\theta=0$ would generate an explicit method).
\end{example}

In general, for the solution of Volterra equations of second kind
($\eta \neq 0$) it is known, that for \emph{any} choice of collocation parameters $0 \leq c_1 < \dots < c_{m+1} \leq 1$ the resulting method is at least of order $O(h) = h^{m+1}$ \cite[Theorem 5.3.2]{Brunner1986}. 
However, the maximum convergence rate can only be achieved if the kernel $G(t, s, y)$ is smooth enough. In the above example of the \gls{srr} model, where a
non-analytic dependence on $|\dot\gamma(t)|$ appears in the model definition,
this implies that the maximum possible order of convergence by these methods is limited by $O(h) = h$ if the exact solution $u$ changes sign (e.g. if $u(t_\ast) = 0$ for some $t_\ast\in[t_0,t]$). The collocation method still works since it only requires the solution to be $C^1$ (once continuously differentiable), which is the case if $\sigma$ is sufficiently smooth. If the exact solution does not change sign numerical approximation schemes with higher orders are possible (see, e.g., Sec.~\ref{ch:application} for the example of creep).

In general, the integrals remaining in the collocation equation will have to
be evaluated numerically. In principle, one can refine separate numerical
grids to solve the integrals on each collocation interval to the maximum
available numerical precision. However, it is more straightforward to
fix a quadrature rule with a given numerical precision. In this latter
case it is common to use the collocation points $t_{i,k}$ as the
abscissas for the quadrature rule \cite{Brunner2004}. We adopt this
practice here.

A strong merit of the collocation ansatz is the flexibility in dealing
with generalized Volterra operators: it leads to a self-consistent ansatz
for elastic moduli of the form $G(t,s,[u])$, where the approximation
replaces that with $G(t,s,[p])$ for each collocation interval.

\begin{example}[Collocation in the \gls{so} model using constant splines]
\label{ex:SO-colloc-const}
As in the previous example we choose the collocation polynomials $p_i(t) = p_i$ to be constant, and we set as collocation parameter $c_1 = \theta \in (0, 1]$. Recall that in the \gls{so} model (Example~\ref{ex:SO-introduction}) the strain-history dependend elastic modulus is of the form
\[
G(t, s, [u]) = G(t, s, {\textstyle\int_s^t u(t') dt'}, u(s))\,.
\]
Inserting the collocation anstatz yields
\begin{equation}
G(t_{i} + \theta h_i, s, [p]) = G\left(t_i + \theta h_i, s, \int_s^{t_i + \theta h_i} p(t') dt', p_j\right) ,\quad s \in[t_j, t_{j+1}]
\end{equation}
Since the integral over the collocation polynomial can be performed exactly,
\begin{equation} \int_s^{t_{i} + \theta h_i} p(t') dt' = \begin{cases}
(t_i + \theta h_i - s) p_i, &\quad s \in [t_i, t_i + \theta h_i]
\\ \theta h_i p_i + \sum_{k = j+1}^{i-1} h_{k} p_{k}   + (t_{j+1} - s) p_j , &\quad s \in [t_j, t_{j+1}]
\end{cases} 
\end{equation}
this leads to the collocation equation
\begin{align}
\sigma(t_{i} + \theta h_i) &= \eta p_i + \int_{t_i}^{t_i+\theta h_i} G\left(t_{i} + \theta h_i, s, (t_{i}+\theta h_i - s) p_i , p_i\right) p_i ds
\\ &\quad+ \sum_{j = 0}^{i-1} \int_{t_{j}}^{t_{j+1}} G\left(t_{i} + \theta h_i, s, \int_s^{t_{i} + \theta h_i} p(t') dt' , p_j\right) p_j ds \nonumber
\end{align}
We fix a quadrature rule for the remaining integrals based on the collocation
parameter, using the rectangular rule with abscissas $t_{j,1}=t_j+\theta h_j$,
\begin{align}
&\int_{t_j}^{t_{j+1}} G\bigg(t_{i} + \theta h_i, s, \int_s^{t_{i} + \theta h_i} p(t') dt' , p_j\bigg) p_j ds \approx h_j p_j G\bigg(t_{i} + \theta h_i, t_{j} + \theta h_j, z_{ij} , p_j\bigg)
\\ &z_{ij} = \int_{t_{j} + \theta h_j}^{t_i + \theta h_i} p(t') dt'
\end{align}
Identifying again $u_{i+1} = p_i$ and rearranging some terms leads to the following method:
\begin{align}
u_{i+1} &= \frac{\sigma(t_i + \theta h_i)}{\eta} + \frac{\theta h_i u_{i+1}}{\eta} G\bigg(t_{i} + \theta h_i, t_{i} + \theta h_i, 0 , u_{i+1} \bigg) 
\\&\quad + \sum_{j = 0}^{i-1} \frac{h_j u_{j+1}}{\eta} G\bigg(t_{i} + \theta h_i, t_{j} + \theta h_j, z_{ij} , u_{j+1} \bigg) \nonumber
\\ z_{ij} &= \theta h_i u_{i+1} + \sum_{k = j+1}^{i-1} h_k u_{k+1} + h_j (1-\theta) u_{j+1}
\end{align}
Note, that the choice $\theta = 1$ leads to a method that could alternatively also have been obtained by direct quadrature (see section \ref{sec:DQ}) with the rightpoint--rectangular rule to approximate $\int G\big(t,s,z(t,s), u(s)\big) u(s) ds$ and $z(t,s) = \int_s^t u(t') dt'$. One expects this method to be of first order in analogy to example \ref{ex:Collocation_SRR}, which we indeed observe in our test case (Sec.~\ref{sec:testcase}).
\end{example}

\begin{table}
\begin{tabular}{c|c|c|c}
Method & Collocation parameters & Convergence order & Quadrature weights
\\
\hline 
Radau~II constant & $c_1 = 1$ & $O(h)$ & $w_1 = 1$
\\ 
\hline 
Lobatto linear & \begin{tabular}{c} $c_1 = 0$ \\ $c_2 = 1$ \end{tabular} & $O(h^2)$ & \begin{tabular}{c} $w_1 = 1/2$ \\ $w_2 = 1/2$ \end{tabular}  
\\ 
\hline 
Radau~II linear & \begin{tabular}{c} $c_1 = 1/3$ \\ $c_2 = 1$ \end{tabular}  & $O(h^3)$ & \begin{tabular}{c} $w_1 = 3/4$ \\ $w_2 = 1/4$ \end{tabular} 
\\ 
\hline 
Lobatto quadratic & \begin{tabular}{c} $c_1 = 0$ \\ $c_2 = 1/2$ \\ $c_3 = 1$ \end{tabular}  & $O(h^4)$ & \begin{tabular}{c} $w_1 = 1/6$ \\ $w_2 = 4/6$ \\ $w_3 = 1/6$ \end{tabular}  
\\ 
\hline 
Radau~II quadratic & \begin{tabular}{c} $c_1 = (4-\sqrt{6})/10$ \\ $c_2 = (4+\sqrt{6})/10$ \\ $c_3 = 1$ \end{tabular}  & $O(h^5)$ & \begin{tabular}{c} $w_1 = (16-\sqrt{6})/36$ \\ $w_2 = (16 + \sqrt{6})/36$ \\ $w_3 = 4/36$ \end{tabular}  
\\ 
\hline 
Lobatto cubic & \begin{tabular}{c} $c_1 = 0$ \\ $c_2 = (5-\sqrt{5})/10$ \\ $c_3 = (5+\sqrt{5})/10$ \\ $c_4 = 1$ \end{tabular}  & $O(h^6)$ & \begin{tabular}{c} $w_1 = 1/12$ \\ $w_2 = 5/12$ \\ $w_3 = 5/12$ \\ $w_4 = 1/12$\end{tabular} 
\\
\hline
\end{tabular}
\caption{Exemplary choices of collocation parameters (nomenclature
follows the Ref.~\cite{Brunner1986}), together with the order of
local superconvergence when applied to Volterra equations, and the
quadrature weights of the corresponding quadrature rules.}
\label{tab:collocation-parameters}
\end{table}

There exist typical choices of collocation parameters, and we list six
common ones in Table~\ref{tab:collocation-parameters} together with their
order of convergence when applied to standard Volterra equations.
Note, that for these choices of collocation parameters the order of convergence (at the grid points $t_i$) is larger than the previously cited minimum order of convergence. 
This phenomenon is known as \textit{local superconvergence} \cite[Section 5.3.3]{Brunner1986}.

\section{Numerical Experiments}\label{sec:numerical_experiments}

\subsection{Test case for experimental convergence analysis}\label{sec:testcase}

We apply different variants of the collocation methods to the
\gls{so} model as a test case
with model parameters $G = 1$, $\eta = 0.1$, $\tau = \infty$, $\gamma_c = 0.1$, $\gamma_\ast = 0.05$ and $\gamma_{\ast\ast} = 0.1$. That model suggests itself because its
structure makes it the numerically most demanding of the examples given
in Sec.~\ref{ch:examples}, while its result are of relevance for the
creep and ``sudden yielding'' of glass-forming colloidal suspensions
\cite{Frahsa2013}.

To obtain error estimates, we first construct an ``exact solution''. To
this end, we impose the strain rate
\begin{equation}
\dot\gamma(t) = u_\text{ref}(t) = \frac{\gamma_c}{\tau_V} \left(1 + 0.5 \sin\left(\frac{2 \pi}{\tau_V} t \right) \right)\,,
\end{equation}
for which we obtain $\sigma(t)$ by direct quadrature with high accuracy.
The collocation
method is then applied to Eq.~\eqref{eq:ice} with Eq.~\eqref{eq:soG},
in order to reconstruct
$\dot\gamma(t)$ with the imposed $\sigma(t)$.
The imposed solution does not change sign, and thus the potential
nonanalyticity of the \gls{so}-model kernel is not limiting the convergence
order of the methods.
We have implemented the methods in the Python programming language,
making heavy use of the standard libraries \texttt{numpy} and
\texttt{scipy}.
The methods all lead to implicit equations for the values of the
desired function $u(t)$ on the grid points; for these implicit equations
a numerical solver (Newton method implementation in the \texttt{scipy.optimize} package) has been used with a pre-set
tolerance of $10^{-14}$. 
We employed equidistant grids with $N+1$ gridpoints:
\begin{align}
t_0 = 0 < \dots < t_i = i\frac{10 \tau_V}{N} < \dots < t_N = 10 \tau_V \quad \text{, where }\tau_V = \frac{\eta}{G} 
\end{align}
With our choice of model parameters we thus have $h = 1/N$.

\begin{figure}
	\centering
	\includegraphics[width = 0.8\textwidth]{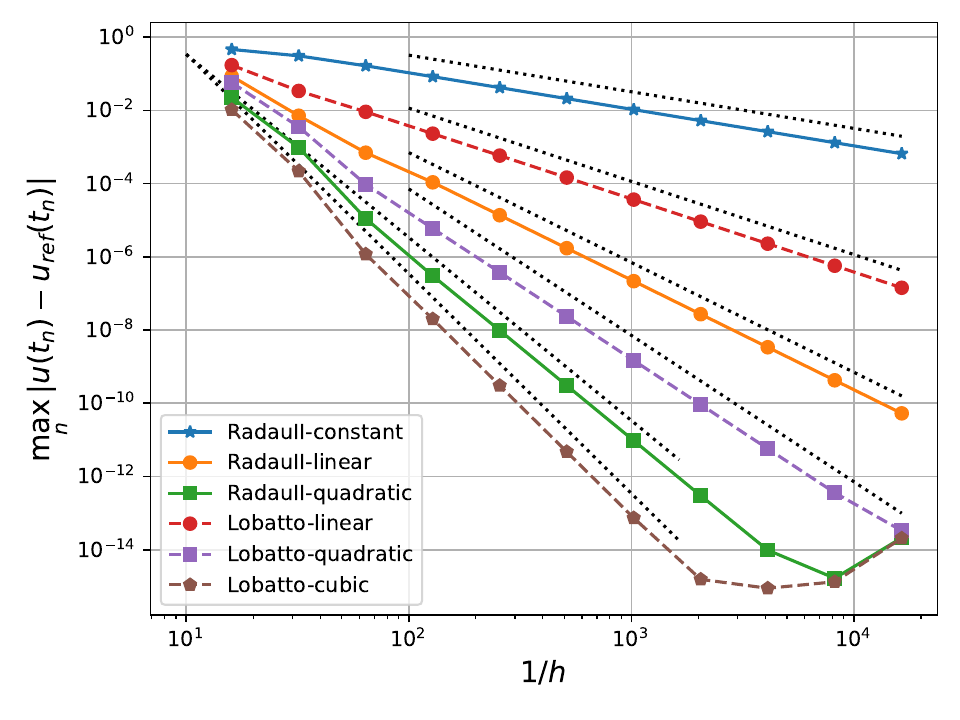} 
	\caption{Convergence rate of different collocation methods applied to the SO-model in the test case described in Sec.~\ref{sec:testcase}. The corresponding collocation parameters are listed in Table~\ref{tab:collocation-parameters}. The dotted lines represent $h^{p}$ with $p = 1, \dots, 6$ respectively.  
	}
	\label{fig:Collocation-u1-SO}
\end{figure} 

In our numerical experiment, we
observe the same convergence rates as for standard Volterra equations (see Fig.~\ref{fig:Collocation-u1-SO}): the Radau~II--constant method (explicitly
written out in Example~\ref{ex:SO-colloc-const})
is of order
$p=1$, while the Lobatto--cubic method reaches $p=6$. Note that in
Fig.~\ref{fig:Collocation-u1-SO} the minimum possible error is
dictated by the accuracy of the implicit equation solver, which explains
the deviations from scaling that appear around an error of $10^{-13}$.

\begin{figure}
	\centering
	\includegraphics[width = 0.8\textwidth]{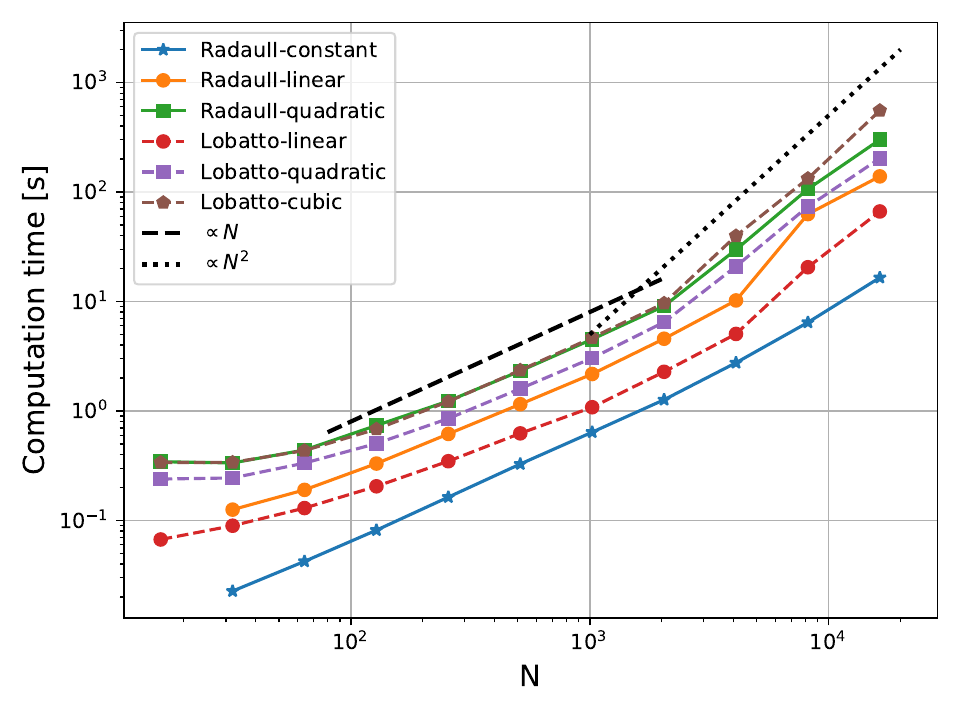}
	\caption{Computational effort of different collocation methods, as in Fig.~\ref{fig:Collocation-u1-SO} with respect to the number of grid points $N$.
	Times are given as total run time of our Python implementation
	on a standard laptop (Intel\textregistered\ Core\texttrademark\ i7-1335U).
	}
	\label{fig:Collocation-u1-SO-b}
\end{figure} 

A usual trade-off in numerics is between accuracy and speed. Note that
in our setting, the asymptotic numerical complexity is determined by the
number of grid points $N$ on the time axis, and due to the convolution
nature of the integral, is expected to scale as $N^2$ for large $N$.
This is indeed seen in Fig.~\ref{fig:Collocation-u1-SO-b} for
$N\gtrsim1000$ in most methods. For smaller $N$ our implementation
shows linear scaling, which we attribute to a large overhead of the
Python code (as opposed to the highly optimized \texttt{numpy} routines)
in this parameter regime.

\begin{figure}
	\centering
	\includegraphics[width = 0.8\textwidth]{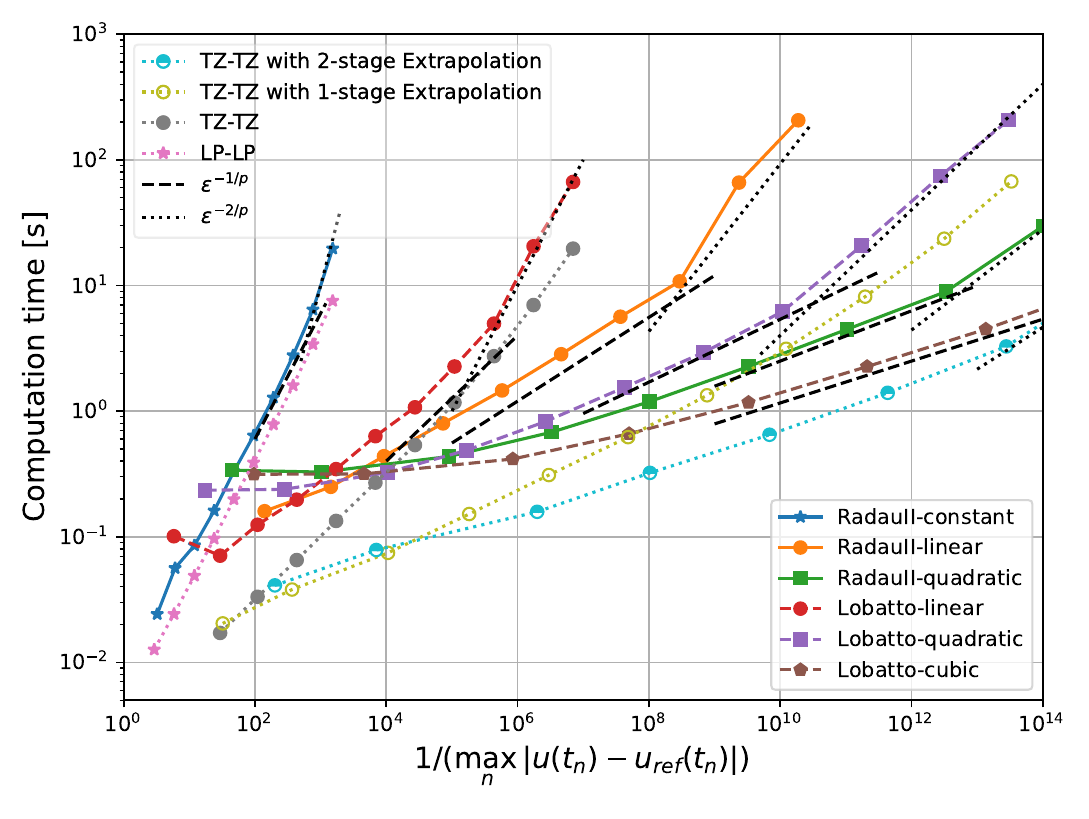}
	\caption{Scaling of the needed computational effort to achieve a desired accuracy $1/\epsilon$ in solving the testproblem for the SO model with different methods described in the text.
}
	\label{fig:Efficiency}
\end{figure} 

To examine which of the implemented methods is the most efficient in solving the testproblem, we show the required computational effort versus the desired accuracy in Fig.~\ref{fig:Efficiency}.
Note that for a method of convergence order $p$, we expect the error
to scale as $\varepsilon\sim h^p$; for a discretization of the time integral
on a linear grid with $N$ points we expect the running time to scale
as $T\sim N^2$ and with $N\sim1/h$ as $T\sim(1/\varepsilon)^{2/p}$.
This is indeed seen in Fig.~\ref{fig:Efficiency} for those cases where the
running time exceeds roughly $10\,\text{s}$; for shorter runs, the
pre-asymptotic implementation details dominate; for the largest $1/\varepsilon$
the finite precision of the floating-point implementation causes deviations.

Note that the trapezoidal--trapezoidal (TZ-TZ) direct quadrature yields a method
equivalent to Lobatto--linear; indeed both show the same scaling, with
differences solely attributed to the difference in our respective
implementations.
Figure~\ref{fig:Efficiency} shows that the Lobatto--cubic method
is extremely efficient for our test case, surpassed only by the
TZ-TZ quadrature with two-stage Richardson extrapolation.
Note that Richardson extrapolation is possible in this test case;
in general for the models with non-analytic dependence on the sought-after
solution, such as the \gls{srr} and \gls{so} model, Richardson extrapolation
cannot be applied.

For applications to physical problems, it is often desirable to cover
large time intervals in the solution; in this case, it is advisable to
choose a method whose error is small even for relatively large step
sizes, as they might be needed in a non-uniform grid. For this reason
especially, the Lobatto--cubic method is an efficient choice in our
setup.

\subsection{Application to Creep}\label{ch:application}

As an exemplary application for our numerical methods, we demonstrate
the creep response of the \gls{so} model. Recall that we define creep
as the response in shear rate to an applied constant stress
$\sigma(t) = \sigma_0 \Theta(t)$ that is instantaneously switched on
at $t=0$.

\begin{figure}
	\centering
	\includegraphics[width = 0.8\textwidth]{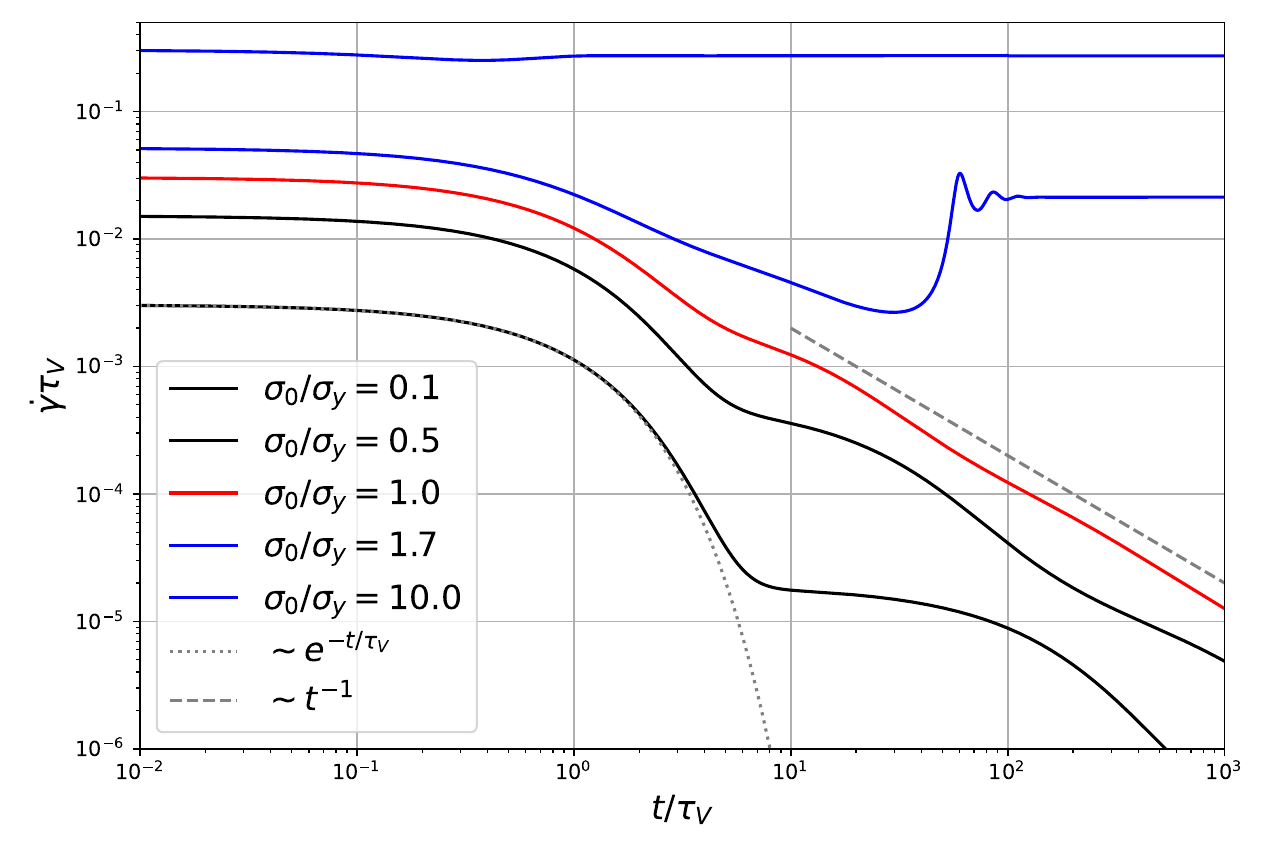}
	\caption{Creep solutions of the stressovershoot model (Example~\ref{ex:SO-introduction}) with exemplary parameters as given in the text,
	for different values of imposed stress $\sigma_0$ in relation
	to the yield stress $\sigma_y$ of the model. The time unit is set
	by the relaxation time of the Kelvin-Voigt model,
	$\tau_V=G_\infty/\eta$. A dashed line denotes the asymptote
	$t^{-1}$ obtained at the yield stress, $\sigma_0=\sigma_y$.}
	\label{fig:Application}
\end{figure} 
Fig.~\ref{fig:Application} shows the numerically obtained solutions for creep in the SO-model for five different applied stresses $\sigma_0$.
For reference, we fix the numerical parameters of the model
as $G_\infty=1$, $\eta=0.1$, $\gamma_c=0.25$, $\gamma_\ast=0.1$, and
$\gamma_{\ast\ast}=0.133$.
This results in a numerically evaluated yield stress of
$\sigma_y \approx 0.03$;
applied stresses are given in relation to this value.

These solutions were obtained on a non-uniform time grid specifically
adapted to a prior estimate of the result, in order to cover a time interval
of up to five orders of magnitude. The grid balances small time steps at short
times where the solution changes rapidly, with larger steps where possible,
reflecting the slow evolution of the solution on exponentially large times.
Specifically, we started by defining a basegrid 
$t^{base}_i=ih_\text{s}$ with $h_\text{s}= 10^{-4} \tau_V$ for $i=0,\ldots 10$;
$t^{base}_i=10^{-3 + (i-10) h_l  }\tau_V $ with $h_l = 0.01$ for $i=11,\ldots 510$;
$t^{base}_i=10^2\tau_V+\big(i-510\big)h_{e}$ with $h_{e} = 0.495 \tau_V$ for $i=511,\ldots 20510$ which we the refined two times by adding points yielding $t_{4 i + k} = t^{base}_i + k (t^{base}_{i+1} - t^{base}_i)/4$, $k = 0, 1, 2, 3$. 
The smallest and largest step lengths thus were $h_\text{min} = t_{40} - t_{41}\approx 0.58\cdot10^{-7}\tau_V$
and
$h_\text{max} = t_{2040}- t_{2039} \approx 0.57\tau_V$, both corresponding to the first and last step in the refined logarithmic part of the grid.\\
For the example, the Lobatto--quadratic method was deemed
sufficiently accurate.
We have checked that the solutions shown in Fig.~\ref{fig:Application} are
also obtained with other variants of the numerical methods described
above, which suggests that they are free from numerical artifacts.

In particular, the numerical method is able to resolve the various qualitatively
different asymptotic regimes for the shear rate $\dot\gamma(t)$ as the
imposed stress $\sigma_0$ is varied across the yield stress.
For $\sigma_0\ll\sigma_y$, the solution is a sequence of increasingly
smeared-out exponential decays, with the initial decay given by the
Kelvin-Voigt relaxation rate.
For $\sigma_0\gg\sigma_y$ on the other hand, the solutions are asymptotically
constant, $\dot\gamma(t)\sim\text{const.}$ for $t\to\infty$; these two
cases represent solid-like and fluid-like response in physical terms.

The region around the yield stress is of particular interest.
For $\sigma_0/\sigma_y\approx1$, an asymptotic decay $\dot\gamma(t)\sim1/t$
is observed (cf.\ dashed line in Fig.~\ref{fig:Application}).
The model thus predicts so-called logarithmic creep (since
$\gamma(t)\sim\ln t$ for $t\to\infty$), i.e., a state in-between the
finitely deformed solid and a stationary flowing liquid.
For $\sigma_0\gtrsim\sigma_y$, a cross-over from the creep-asymptote
to constant flow is observed, and due to the presence of the
stress overshoot in the model (i.e., a non-monotonic, not strictly
positive function $G$), there is a relatively rapid crossover
that also involves a non-monotonic time dependence of the shear rate.
The rapid crossover is also termed ``sudden yielding'' in the physics
literature and in particular associated to the sudden failure of
a large class of glasses and gels (see, e.g., Ref.~\cite[Fig.~1]{Aime2018}).
We postpone a more in-depth discussion of the features of this model
to a future publication.

\section{Conclusion}\label{sec:discussion}

\glsresetall
We developed a numerical scheme to solve common \glspl{ice} of
nonlinear rheology that are formulated as generalized Volterra-type
integral equations, Eq.~\eqref{eq:ice}.
In particular, we transferred standard collocation methods that have
been developed for classical Volterra integral equations of second kind,
and demonstrate their convergence when applied to a set of
physically motivated examples. While simpler direct-quadrature methods
can give similar performance when combined with Richardson extrapolation
(where allowed by the structure of the integral kernel), for the case
of the collocation methods we are able to prove convergence to the
unqiue solution of the problem.
The proof is an adaptation of the techniques used to prove uniqueness,
existence of, and convergence to the solutions in the case of ordinary
Volterra integral equations.

Physically, the models we discuss here are of interest when the
entire strain-rate history enters the nonlinear shear modulus, and one
seeks a a solution to the stress-controlled rheology problem, i.e.,
to determine the strain and strain rates resulting from an applied
stress. As a particular example we demonstrate the applicability of our
collocation method to determine the creep-response of a glass-forming
model that incorporates stress overshoots and with it, ``sudden yielding''
phenomena.

Although not the center of interest in the present application,
our proofs extend also to certain \glspl{avie},
Eq.~\eqref{eq:avie}, as long as a finite time interval is considered
and a continuous solution exists. In this regard, we extend the global
convergence proof of \Citeauthor{Zhang.2016} \cite{Zhang.2016}, although
we do not prove superconvergence of the collocation method; we also do
not prove existence of a unique continuous solution here. Some comments
can be found in Appendix~\ref{app:convergence-proof}.

For now, the methods that we presented are all ``onestep'' methods.
Multi-step methods might be required in cases where the solutions are
sought over extremely large time intervals and still might involve
rapid changes at late times. The error analysis of multi-step methods
is more intricate and has to be left for future work. For now, the salient
assumption is that in many physical problems, the solutions will
be slowly varying functions, or moderate deviations from that (as in
the case of ``sudden'' yielding). One relevant application for improved
multi-step methods might be given by large-amplitude oscillatory tests,
where the imposed stress $\sigma(t)\propto\sin(\omega t)$ imposes
a time scale $1/\omega$ that needs to be honored by the numerical grid
at all times.

With the groundwork laid down here, a future task would be to apply the methods to more intricate models as the schematic ITT-MCT model \cite{Brader2009}.
There, the integral constitutive equation \eqref{eq:ice} is complemented
by a closure that relates the kernel $G(t,t',[\dot\gamma])$ to dynamical
density correlation functions $\phi(t,t',[\dot\gamma])$ that in turn
are solutions of a set of nonlinear Volterra equations of autoconvolution
type for $u_{t'}(t)=\phi(t,t',[\dot\gamma])$-- with additional parametric dependence on $t'$.
While the methods developed here will still be applicable, the solution of
such a one-parameter family of equations together with an appropriate choice
of $(t,t')$ grid to cover exponentially large times and time differences,
is numerically extremely demanding.

\subsection*{Acknowledgements}

We thank J\"urgen Saal for continued valuable input in the preparation
of the analytical background.

\subsection*{Data availability}

Python code to implement the collocation methods discussed here,
a Jupyter notebook demonstrating the calculation of creep response for
the stress-overshoot model, and the data corresponding to Fig.~\ref{fig:Application}
are publicly available as a Zenodo repository \cite{zenodo}.

\subsection*{Funding}
This research did not receive any specific grant from funding agencies in the public, commercial, or not-for-profit sectors.

\subsection*{CRediT authorship contribution statement}
\textbf{Stephan Domann:} Writing - Original Draft, Investigation, Software. \textbf{Thomas Voigtmann:} Writing - Review \& Editing, Supervision, Software.

\bibliographystyle{apsrev4-2} 
\bibliography{numerics} 

@Article{Amat2023,
  author  = {Amat, Sergio and Li, Zhilin and Ruiz-{\'A}lvarez, Juan and Solano, Concepci{\'o}n and Trillo, Juan C},
  journal = {Math. Comput. Sim.},
  title   = {Numerical integration rules with improved accuracy close to discontinuities},
  year    = {2023},
  pages   = {593--614},
  volume  = {210},
  doi     = {10.1016/j.matcom.2023.03.032},
}

@Article{Bonfanti2020,
  author  = {Bonfanti, Alessandra and Kaplan, Jonathan Louis and Charras, Guillaume and Kabla, Alexandre},
  journal = {Soft Matter},
  title   = {Fractional viscoelastic models for power-law materials},
  year    = {2020},
  number  = {26},
  pages   = {6002--6020},
  volume  = {16},
  doi     = {10.1039/d0sm00354a},
}

@Book{Gripenberg1990,
  author    = {Gripenberg, Gustaf and Londen, Stig-Olof and Staffans, Olof},
  publisher = {Cambridge University Press},
  title     = {Volterra integral and functional equations},
  year      = {1990},
  isbn      = {9780521103060},
}

@Book{Brunner1986,
  author = {Brunner, Hermann and van der Houwen, Pieter Jacobus},
  title  = {The numerical solution of {V}olterra equations},
  publisher = {Elsevier},
  address = {Amsterdam},
  year   = {1986},
}

@Article{Amann2013,
  author  = {Amann, Christian P and Siebenb{\"u}rger, Miriam and Kr{\"u}ger, Matthias and Weysser, Fabian and Ballauff, Matthias and Fuchs, Matthias},
  journal = {J. Rheol.},
  title   = {Overshoots in stress-strain curves: Colloid experiments and schematic mode coupling theory},
  year    = {2013},
  number  = {1},
  pages   = {149--175},
  volume  = {57},
  doi     = {10.1122/1.4764000},
}

@Article{Frahsa2013,
  author  = {Frahsa, Fabian and Bhattacharjee, Amit Kumar and Horbach, J{\"u}rgen and Fuchs, Matthias and Voigtmann, Thomas},
  journal = {J. Chem. Phys.},
  title   = {On the Bauschinger effect in supercooled melts under shear: Results from mode coupling theory and molecular dynamics simulations},
  year    = {2013},
  pages   = {12A513},
  volume  = {138},
  doi     = {10.1063/1.4770336},
}

@Article{Bogue1966,
  author  = {Bogue, DC},
  journal = {Ind. Engin. Chem. Fundam.},
  title   = {Explicit constitutive equation based on integrated strain history},
  year    = {1966},
  number  = {2},
  pages   = {253--259},
  volume  = {5},
  doi     = {10.1021/i160018a016},
}

@Article{Siebenbuerger2012,
  author  = {Siebenb{\"u}rger, Miriam and Ballauff, Matthias and Voigtmann, Th},
  journal = {Phys. Rev. Lett.},
  title   = {Creep in colloidal glasses},
  year    = {2012},
  number  = {25},
  volume  = {108},
  doi     = {10.1103/physrevlett.108.255701},
}

@Book{Wazwaz2011,
  author    = {Wazwaz, Abdul-Majid},
  publisher = {Springer},
  title     = {Linear and nonlinear integral equations},
  year      = {2011},
  doi       = {10.1007/978-3-642-21449-3},
}

@Book{Hackbusch1995,
  author    = {Hackbusch, Wolfgang},
  publisher = {Birkh{\"a}user},
  title     = {Integral equations: theory and numerical treatment},
  year      = {1995},
}

@Book{Brunner2004,
  author    = {Brunner, Hermann},
  publisher = {Cambridge university press},
  title     = {Collocation methods for {V}olterra integral and related functional differential equations},
  year      = {2004},
  isbn      = {9780511261992},
}

@Book{Linz1985,
  author    = {Linz, Peter},
  publisher = {SIAM},
  title     = {Analytical and numerical methods for {V}olterra equations},
  year      = {1985},
}

@Article{Wolkenfelt1981,
  author  = {Wolkenfelt, PHM},
  journal = {BIT Numer. Math.},
  title   = {Reducible quadrature methods for {V}olterra integral equations of the first kind},
  year    = {1981},
  number  = {2},
  pages   = {232--241},
  volume  = {21},
  doi     = {10.1007/bf01933168},
}

@Book{Stoer2002,
  author    = {Stoer, Josef and Bulirsch, Roland and Bartels, R and Gautschi, Walter and Witzgall, Christoph},
  publisher = {Springer},
  title     = {Introduction to numerical analysis},
  year      = {2002},
  doi       = {10.1007/978-0-387-21738-3},
}

@Article{Brader2009,
  author  = {Brader, Joseph M and Voigtmann, Thomas and Fuchs, Matthias and Larson, Ronald G and Cates, Michael E},
  journal = {Proc. Natl. Acad. Sci. (USA)},
  title   = {Glass rheology: From mode-coupling theory to a dynamical yield criterion},
  year    = {2009},
  number  = {36},
  pages   = {15186--15191},
  volume  = {106},
  doi     = {10.1073/pnas.0905330106},
}

@Article{FuchsBraderPRL,
  author   = {J. M. {Brader} and M. E. {Cates} and M. {Fuchs}},
  journal  = {Phys.~Rev.~Lett.},
  title    = {First-Principles Constitutive Equation for Suspension Rheology},
  year     = {2008},
  pages    = {138301},
  volume   = {101},
  doi      = {10.1103/physreve.86.021403},
  numpages = {4},
}

@Article{Fuchs2005a,
  author  = {Fuchs, Matthias and Ballauff, Matthias},
  journal = {J. Chem. Phys.},
  title   = {Flow curves of dense colloidal dispersions: Schematic model analysis of the shear-dependent viscosity near the colloidal glass transition},
  year    = {2005},
  number  = {9},
  volume  = {122},
  doi     = {10.1063/1.1859285},
}

@Article{Siebenbuerger2009,
  author  = {Siebenb{\"u}rger, Miriam and Fuchs, Matthias and Winter, Henning and Ballauff, Matthias},
  journal = {J. Rheol.},
  title   = {Viscoelasticity and shear flow of concentrated, noncrystallizing colloidal suspensions: Comparison with mode-coupling theory},
  year    = {2009},
  number  = {3},
  pages   = {707--726},
  volume  = {53},
  doi     = {10.1122/1.3093088},
}

@Article{Zausch2008,
  author  = {Zausch, Jochen and Horbach, J{\"u}rgen and Laurati, Marco and Egelhaaf, Stefan Ulrich and Brader, Joseph M and Voigtmann, Th and Fuchs, Matthias},
  journal = {J. Phys.: Cond. Matt.},
  title   = {From equilibrium to steady state: the transient dynamics of colloidal liquids under shear},
  year    = {2008},
  number  = {40},
  pages   = {404210},
  volume  = {20},
  doi     = {10.1088/0953-8984/20/40/404210},
}

@Article{Aime2018,
  author  = {Aime, Stefano and Ramos, Laurence and Cipelletti, Luca},
  journal = {Proc. Natl. Acad. Sci. (USA)},
  title   = {Microscopic dynamics and failure precursors of a gel under mechanical load},
  year    = {2018},
  number  = {14},
  pages   = {3587--3592},
  volume  = {115},
  doi     = {10.1073/pnas.1717403115},
}

@Book{Kubrusly2026,
  author    = {Kubrusly, Carlos S},
  publisher = {Springer},
  title     = {Essentials of measure theory},
  year      = {2026},
  edition   = {2},
  doi       = {10.1007/978-3-032-12663-4},
}

@Book{Amann2008,
  author     = {Amann, Herbert and Escher, Joachim},
  publisher  = {Birkhäuser},
  title      = {Analysis {II}},
  year       = {2008},
  doi        = {10.1007/978-3-7643-7480-8},
  translator = {Levy, Silvio and Cargo, Matthew},
}

@Article{Maxwell.1867,
  author  = {J. Clerk {Maxwell}},
  journal = {Philos.~Trans.~R.~Soc.~London},
  title   = {On the Dynamical Theory of Gases},
  year    = {1867},
  pages   = {49--88},
  volume  = {157},
}

@Article{Voigtmann.2012,
  author  = {Th. {Voigtmann} and J. M. {Brader} and M. {Fuchs} and M. E. {Cates}},
  journal = {Soft Matter},
  title   = {Schematic mode coupling theory of glass rheology: single and double step strains},
  year    = {2012},
  pages   = {4244--4253},
  volume  = {8},
  doi     = {10.1039/C2SM06891E},
}

@Article{Caraglio.2020,
  author  = {Michele {Caraglio} and Lukas {Schrack} and Gerhard {Jung} and Thomas {Franosch}},
  journal = {Commun. Comput. Phys.},
  title   = {An Improved Integration Scheme for Mode-Coupling-Theory Equations},
  year    = {2020},
  pages   = {628--648},
  volume  = {29},
  doi     = {10.4208/cicp.OA-2020-0125},
}

@Article{Tuck.1967,
  author  = {E. O. {Tuck}},
  journal = {Math.~Comp.},
  title   = {A Simple {\textquotedblleft}Filon-Trapezoidal{\textquotedblright} Rule},
  year    = {1967},
  pages   = {239--241},
  volume  = {21},
  doi     = {10.1090/s0025-5718-67-99892-4},
}

@Article{Zhang.2016,
  author  = {Ran {Zhang} and Hui {Liang} and Hermann {Brunner}},
  journal = {SIAM J.~Num.~Analysis},
  title   = {Analysis of Collocation Methods for Generalized Auto-Convolution Volterra Integral Equations},
  year    = {2016},
  pages   = {899--920},
  volume  = {54},
  doi     = {10.1137/15M1019362},
}

@Book{Kaye1962,
  author    = {Kaye, A},
  publisher = {College of Aeronautics},
  title     = {Non-{N}ewtonian flow in incompressible fluids},
  year      = {1962},
}

@article{Suzuki.2013,
  author = {Koshiro {Suzuki}  and Hisao {Hayakawa} },
  title = {Nonequilibrium mode-coupling theory for uniformly sheared
    underdamped systems},
  journal = {Phys.~Rev.~E},
  volume = {87},
  year = {2013},
  pages = {012304},
  numpages = {27},
  doi = {10.1103/PhysRevE.87.012304},
}

@article{Chong.2010,
  author = {Song-Ho {Chong}  and Michio {Otsuki}  and Hisao {Hayakawa} },
  title = {Generalized Green-Kubo relation and integral fluctuation theorem for
    driven dissipative systems without microscopic time reversibility},
  journal = {Phys.~Rev.~E},
  volume = {81},
  year = {2010},
  pages = {041130},
  numpages = {4},
  doi = {10.1103/PhysRevE.81.041130},
}

@Article{Fuchs.2003,
  author  = {Matthias {Fuchs} and Michael E. {Cates}},
  journal = {Faraday~Discuss.},
  title   = {Schematic models for dynamic yielding of sheared colloidal glasses},
  year    = {2003},
  pages   = {267--286},
  volume  = {123},
  doi     = {10.1039/b205629a},
}

@article{Goetze.1995b,
  author = {W. {G\"otze}  and L. {Sj\"ogren} },
  title = {General Properties of Certain Non-linear Integro-Differential
    Equations},
  journal = {J.~Math.~Analysis and Appl.},
  volume = {195},
  year = {1995},
  pages = {230--250},
  doi = {10.1006/jmaa.1995.1352},
}

@article{Fuchs.2002c,
  author = {M. {Fuchs}  and M. E. {Cates} },
  title = {Theory of Nonlinear Rheology and Yielding of
    Dense Colloidal Suspensions},
  journal = {Phys.~Rev.~Lett.},
  volume = {89},
  year = {2002},
  pages = {248304},
  numpages = {4},
  doi = {10.1103/PhysRevLett.89.248304},
}

@article{Fuchs.2009,
  author = {M. {Fuchs}  and M. E. {Cates} },
  title = {A mode coupling theory for Brownian particles in homogeneous
    steady shear flow},
  journal = {J.~Rheol.},
  volume = {53},
  year = {2009},
  pages = {957--1000},
  doi = {10.1122/1.3119084},
}

@Misc{zenodo,
  author    = {Stephan Domann and Thomas Voigtmann},
  note      = {{DOI:}10.5281/zenodo.22978794},
  title     = {Collocation solvers for creep response of integral constitutive equations},
  year      = {2026},
  doi       = {10.5281/zenodo.22978794},
  publisher = {Zenodo},
  url       = {https://doi.org/10.5281/zenodo.22978794},
}

\begin{appendix}

\section{Existence and uniqueness of solutions}

\subsection{Nonlinear Volterra equations and the SRR model}\label{app:nonlinear-Volterra}
In the following let $I = [t_0, T] \subset \mathbb{R}$ or $I = [t_0, \infty )$ be the time-interval on which we want to know the solution $u$. Define $D = \{(t, s) \in I \times I; s \leq t \}$. 
For nonlinear Volterra equations of second kind
\begin{align}
u(t) = f(t) + \int_{t_0}^t k(t, s, u(s) ) ds \label{equ:app_Volterra}
\end{align}
one can find the following theorem \cite{Hackbusch1995}[Ch. 2.1.1]:
\begin{theorem}
Assume $f$ and $k$ to be continuous functions ($f \in C(I)$ and $k \in C(D \times \mathbb{R})$). Furthermore let the kernel $k$ fulfill the following Lipschitz condition: There is a function $L \in C(D)$ such that
\begin{align}
|k(t, s, x) - k(t, s, y)| \leq L(t, s) |x - y| \quad \text{for all } (t, s)\in D; \, x,y \in \mathbb{R}  
\end{align}
Then, there is exactly one solution $u \in C(I)$ of the integral equation \eqref{equ:app_Volterra}. 
\end{theorem}

\begin{example}[Strainrate relaxation model]
In the case of the \gls{srr} model (Example \ref{ex:SRR_introduction} with
$\eta \neq 0$) the forcing function $f$ is given by $f(t) = \sigma(t)/\eta$ and the kernel $k$ is given by
\begin{align}
k(t, s, y) = \frac{G_\infty}{\eta} e^{-\frac{t-s}{\tau}} e^{-\frac{(t-s)|y|}{\gamma_c}} y
\end{align}
All we need to do is to check the Lipschitz-type condition for $k(t, s, y)$:
\begin{align}
|k(t, s, x) - k(t, s, y)| &= \frac{G_\infty}{\eta} e^{-\frac{t-s}{\tau}} \bigg|e^{-\frac{(t-s)|x|}{\gamma_c}} x - e^{-\frac{(t-s)|y|}{\gamma_c}} y \bigg| \label{equ:app_SRR-lip}
\end{align} 
But $h(x)=xe^{-\alpha|x|}$, $\alpha\geq0$, is Lipschitz; as can be observed from
the mean-value theorem in integral form
\begin{align}
|h(x) - h(y)| \leq \int_{x}^{y} |h'(r)| dr \leq |x-y| \sup_{r \in \mathbb{R}\setminus \{0\}} |h'(r)|  = |x-y|\,,
\end{align}
i.e., a Lipschitz condition with $L=1$.
Inserting this result into Eq.~\eqref{equ:app_SRR-lip} yields that the \gls{srr} model has a unique solution.

\end{example}

\subsection{General functional equations and the stressovershoot model}\label{app:functional-equations}

Due to the more abstract viewpoint of functional equations the existence and uniqueness theorem we cite here is more intricate and requires substantially more effort to check. 
Before citing the theorem itself some introduction of notation and explanatory comments are in order.

We want to find a solution of the functional equation given by

\begin{align}
u(t) = H[u](t)\,, \quad t \in [t_0, T_{max}) \subset \mathbb{R} \label{equ:app_functionalequ}
\end{align}
where $H$ is some nonlinear operator and where $T_{max}$ is the a priori unknown maximum time of existence.
This is a fixpoint equation. 
For our purposes we are interested in continous solutions, so we need to $H$ to be an operator that maps a certain set of continous functions to a continous function, e.g. $\phi \mapsto H[\phi] \in C([t_0, T]; \mathbb{R}^n)$ for $\phi \in A \subset C([t_0, T]; \mathbb{R}^n)$
for all $T < T_{max}$. To have a notion of continuity of the nonlinear operator $H: A \subset C([t_0, T]; \mathbb{R}^n) \rightarrow C([t_0, T]; \mathbb{R}^n)$ we need to fix a norm on $C([t_0, T]; \mathbb{R}^n)$. As norm we use the supremum norm
\begin{align}
\lVert \phi \rVert_\infty = \sup_{t \in [t_0, T]} | \phi(t) |
\end{align}
where $|x|, x \in \mathbb{R}^n$ denotes the norm in $\mathbb{R}^n$. To ensure the pointwise continuity of the operator $H$ will be the role of condition \textit{1)} in the following theorem.
 
The solution $u$ defines curves $(t, u(t))_{t \in [t_0, T]}$ in $\mathbb{R}_{\geq t_0} \times \mathbb{R}^n$, $T < T_{max}$. 
To capture cases where $H$ is not defined for functions $\phi$ that exceed a certain bound or where $H$ is not defined for some time-instances $t \geq T_H$, we have to  restrict ourselves to functions $\phi$ where the induced curves $(t, \phi(t))_{t \in [t_0, T]}$ lie within some open, connected subset $M \subseteq \mathbb{R}_{\geq t_0} \times \mathbb{R}^n$. This $M$ has to be adjusted according to the operator $H$ at hand. 
Thus, the function space where we search for a solution is given by
\begin{align}
A = C_{\xi,M}([t_0, T]; \mathbb{R}^n) = \bigg\{ \phi \in C([t_0, T]; \mathbb{R}^n) \bigg| \phi(t_0) = \xi, (t, \phi(t)) \in M \text{ for } t \in [t_0, T] \bigg\} .
\end{align}
The choice of $M$ naturally defines an upper bound for the length of the time-intervals $[t_0, T]$ given by
\begin{align}
T^M_\infty = \sup\bigg\{ T \in \mathbb{R}_{> t_0} \bigg| C_{\xi, M}([t_0, T], \mathbb{R}^n) \neq \emptyset \bigg\}
\end{align}
Note, $T^M_\infty = \infty$ is allowed.

We resort to the Banach fixpoint theorem to establish uniqueness of the
solution of the functional equation \eqref{equ:app_functionalequ}.
This will be the origin of condition \textit{4)} in the theorem, where we in principle demand that $H$ defines a contraction on some chosen subset of $C_{\xi, M}([0, T], \mathbb{R}^n)$. 

Since we cannot expect $H$ to be a contraction on the whole space $C_{\xi, M}([t_0, T];\mathbb{R}^n)$ we restrict this condition to functions that are equal up to some time-instance $T_\text{eq} < T$ which is sufficiently close to $T$. The full unique solution is then constructed by starting to construct a solution on a small time-interval $[t_0, T_1]$ and then subsequently constructing continuations on the time-intervalls $[T_i, T_{i+1}]$. This is the origin of the assumptions \textit{4a)} and \textit{4b)} in the theorem below.
A possibly infinite sequence of such steps can assert uniqueness
on the right-open interval $[t_0,T_\text{max}]$.
In the case $T_\text{max}<T_\infty^M$, one can show with some additional
hypotheses that then the solution $u$ has to either diverge
($|u(t)| \rightarrow \infty$ as $t\uparrow T_{max}$)
or approach the boundary of $M$
(e.g. $\big( T_{max}, u(t\uparrow T_{max}) \big) \in \partial M \,$ ).
The latter case may appear if the curve $(t, u(t))$, $t \in [t_0, T_{max})$, takes values such that $H[u](T_{max})$ is not defined anymore. 

With these qualifications, the following theorem given by \citeauthor{Gripenberg1990} \cite[Ch. 12 Theorem 2.1 and Corollary 2.2]{Gripenberg1990} applies (with minor change in notation):
\begin{theorem}\label{theorem:app_ex2}
Assume that the following conditions are satisfied:
\begin{enumerate}[label=\arabic*)]
\item For each $T \in [t_0, T^M_\infty)$, $H$ maps $C_{\xi, M}([t_0, T];\mathbb{R}^n)$ into $C([t_0, T]; \mathbb{R}^n)$ and for each $t \in [t_0, T]$ the mapping $\phi \mapsto H[\phi](t)$, $\phi \in C_{\xi, M}([t_0, T]; \mathbb{R}^n)$, is continous with respect to the supremum norm $\lVert \cdot \rVert_\infty$. 
\item $H$ is causal, e.g. if $\tilde{t} \in [t_0, T)$ and if $\phi(t) = \psi(t)$ for $t \leq \tilde{t}$ then $H[\phi](t) = H[\psi](t)$. 
\item $H[\xi](t_0) = \xi$ (initial value compatibility)
\item For each compact connected subset $K \subset M$ containing the starting point $(t_0, \xi)$ there exists constants $\delta_K > 0$ and $\gamma_K \in (0, 1)$ such that the following impliciation holds:
 if for some functions $\phi, \psi \in C_{\xi, K}([t_0, T]; \mathbb{R}^n)$ we have
\begin{enumerate}
\item $\psi(s) = \phi(s)$ for $s \in [t_0, T_{eq} ]$ with $T_{eq} = \textnormal{max}(t_0, T - \delta_K)$
\item $|\psi(s) - \phi(s)|  \leq \delta_K$ for $s \in [T_{eq}, T]$
\end{enumerate} 
then there holds
\begin{align}
\sup_{s \in [T_{eq}, T]} |H[\phi](s) - H[\psi](s)| \leq \gamma_K \sup_{s \in [T_{eq}, T]} |\psi(s) - \phi(s)|
\end{align}
\end{enumerate}
Then there exists a \emph{unique} noncontinuable solution $u$ of Eq.~\eqref{equ:app_functionalequ} fulfilling the initial condition enforced by $H$. The maximal interval of existence is open to the right. e.g. it is given by $[t_0, T_{max})$ for some $T_{max} > t_0$. 
Moreover, if $\gamma_K \leq \frac{1}{2}$ and if the implication 4) is true without the assumption b), then the curve $(t, u(t))$ leaves every compact subset of $M$ as $t \uparrow T_{max}$. 
\end{theorem}
\begin{example}[Stressovershoot model]
The \gls{so} model can be written as
\begin{align}
&\eta u(t) = \sigma(t) - \int_{t_0}^t G_\infty \bigg(1 - \big( \frac{z(t,s, [u])}{\gamma_{\ast}}\big)^{2n} \bigg) e^{- ( \frac{z(t, s, [u])}{\gamma_{\ast \ast}} )^{2n} }  e^{-\frac{t-s}{\tau}} e^{-\frac{(t - s)|u(s)|}{\gamma_c} } u(s) ds \\
&G(t, s, [u]) = G_\infty \bigg(1 - \big( \frac{z(t,s, [u])}{\gamma_{\ast}}\big)^{2n} \bigg) e^{- ( \frac{z(t, s, [u])}{\gamma_{\ast \ast}} )^{2n} }  e^{-\frac{t-s}{\tau}} e^{-\frac{(t - s)|u(s)|}{\gamma_c} } \\
&z(t, t', [u]) = \int_{t'}^t u(s) ds
\end{align}
Assuming $\eta > 0$ we can check the conditions of theorem \ref{theorem:app_ex2} for
\begin{align}
H[u](t) = \frac{\sigma(t)}{\eta} - \int_{t_0}^t \frac{G_\infty}{\eta} \bigg(1 - \big( \frac{z(t,s, [u])}{\gamma_{\ast}}\big)^{2n} \bigg) e^{- ( \frac{z(t, s, [u])}{\gamma_{\ast \ast}} )^{2n} }  e^{-\frac{t-s}{\tau}} e^{-\frac{(t - s)|u(s)|}{\gamma_c} } u(s) ds
\end{align}
The initial condition $\xi$ fixed by $H$ is $\xi = u(t_0) = \sigma(t_0)/\eta$ and one can directly see that $H$ is causal. Functions in the domain of $H$ do not need to be bounded in any fashion,so we can choose $M = \mathbb{R}_{\geq t_0} \times \mathbb{R}$, such that $T^M_{\infty} = \infty$. Lets start by checking condition 1.) of the theorem and fix a $T \in [t_0, \infty)$. 
As long as $\sigma(t)$ is continuous, $H$ will map continous functions $\phi \in C_{\xi, M}([t_0, T])$ to continuous functions on $[t_0, T]$. For any fixed $t \in [t_0, T]$ the continuity of the mapping $\phi \mapsto H[\phi](t)$, $\phi \in C_{\xi, M}([t_0, T])$, is given since this mapping can be seen as a product and chain of continous operations.

This leaves checking whether condition 4.) holds:
Let $K \subset M$ be compact and connected with $(t_0, \xi) \in K$. For later usage we define
\begin{align}
\alpha_K = \max_{(t, x) \in K} |x|
\end{align}
In the following we will determine $\delta_K > 0$ such that $\gamma_K = \frac{1}{3}$ is possible. Take two functions $\phi, \psi \in C_{\xi, K}([0, T])$ that fulfill condition 4a). For $t \in [T_{eq}, T]$ we have
\begin{align}
|H[\phi](t) - H[\psi](t)| &= \bigg| \int_{t_0}^t G(t, s, [\phi]) \phi(s) - G(t, s, [\psi]) \psi(s)  ds \bigg| \\
&\leq \int_{t_0}^t \big| G(t, s, [\phi]) \big| \big| \phi(s) - \psi(s) \big| + \big|\psi(s) \big| \big| G(t, s, [\phi]) - G(t, s, [\psi]) \big| ds \\
&\leq (T - T_{eq}) \max_{t, s, \phi}|G(t, s, [\phi])| \lVert \phi - \psi \rVert_{\infty} \label{equ:app_example-1} \\
&\quad + \int_{t_0}^t \alpha_K \big| G(t, s, [\phi]) - G(t, s, [\psi]) \big| ds \nonumber
\end{align}
Note, that 
\begin{align}
\max_{t, s, \phi}|G(t, s, [\phi])| &= \max\bigg\{ |G(t, s, [\phi])| \bigg| s\leq t;\, t,s \in [t_0, T] \text{ and } \phi \in C_{\xi, K}([0, T]) \bigg\}
< \infty
\end{align}
is finite. 
Next, we have to deal with the second term. Observe that $G(t, s, [\phi])$ is of the form
\begin{align}
G(t, s, [\phi]) = G\big(t, s, \phi(s), z(t, s, [\phi])\big)
\end{align}
where $G(t, s, x, z)$ is Lipschitz-continous both in $x$ and $z$:
\begin{align}
&|G(t, s, x , z ) - G(t, s, x', z')| \leq L_1(t, s) |x - x'| + L_2(t, s) |z - z'| \\
&L_1(t, s) = \frac{G}{\eta} e^{-\frac{t-s}{\tau}} \sup_{x \in \mathbb{R}^n} e^{-\frac{(t-s) |x|}{\gamma_c}} = \frac{G}{\eta} e^{-\frac{t-s}{\tau}}
\\ &L_2(t, s) = \frac{G}{\eta} e^{-\frac{t-s}{\tau}} \sup_{z \in \mathbb{R}^n} |(1 - (z/\gamma_{\ast})^{2n}) e^{- (\frac{z}{\gamma_{\ast\ast}})^{2n}}| .
\end{align}
Thus, when we can estimate $|z - z'| = |z(t, s, [\phi]) - z(t, s, [\psi])|$ with respect to $\delta_K$ or $T-T_{eq}$, we have everything we need to finish the proof. 
\begin{align}
|z(t, s, [\phi]) - z(t, s, [\psi])| &= | \int_{s}^t \phi(t') - \psi(t') dt' |
\\ &\leq (T - T_{eq}) \lVert \phi - \psi \rVert_\infty
\end{align}
Thus, we have (defining $L_i = \max_{s\leq t} L_i(t, s)$)
\begin{align}
\int_{t_0}^t \big| G(t, s, [\phi]) - G(t, s, [\psi]) \big| ds &\leq \int_{t_0}^t \big(L_1(t, s) |\phi(s) - \psi(s)| + L_2(t, s) |z(t, s, [\phi]) - z(t, s, [\psi])| \big) ds
\\ &\leq L_1 (T-T_{eq}) \lVert \phi - \psi \rVert_\infty + L_2 (T-t_0) (T-T_{eq}) \lVert \phi - \psi \rVert_\infty
\end{align}
Going back to equation \ref{equ:app_example-1} we have shown
\begin{align}
|H[\phi](t) - H[\psi](t)| &\leq (T-T_{eq}) 
\bigg( \max_{t, s, \phi}|G(t, s, [\phi])| + L_1 \alpha_K + L_2 \alpha_K (T-t_0)\bigg)
\lVert \phi - \psi \rVert_\infty
\\&\leq \delta_K \bigg( \max_{t, s, \phi}|G(t, s, [\phi])| + L_1 \alpha_K + L_2 \alpha_K (T^K_{\infty}-t_0)\bigg)
\lVert \phi - \psi \rVert_\infty
\end{align}
Thus, we have $|H[\phi](t) - H[\psi](t)| \leq \frac{1}{3}\lVert \phi - \psi \rVert_\infty$ with 
\begin{align}
\delta_K = \frac{1}{3 \bigg( \max_{t, s, \phi}|G(t, s, [\phi])| + L_1 \alpha_K + L_2 \alpha_K (T^K_{\infty}-t_0)\bigg)}
\end{align}
Thus, the stressovershoot model fulfills all conditions of theorem \ref{theorem:app_ex2} and has a unique solution. Furthermore it fullfills the additional conditions in the theorem.
Therefore, if the solution $u(t)$ stays bounded, we have $T_{max} = T^M_\infty = \infty$. 
\end{example}
In the essential steps in the proof of the example the specific form of $G(t, s, x, z)$ was not of importance. We only needed the Lipschitz-type condition for $G(t, s, x, z)$ in both components and the specific estimate for $|z(t,s, [\phi]) - z(t, s, [\psi])|$. Thus, we also have proven:
\begin{proposition}\label{app:prop-stressovershoot}
Let $\sigma(t)$ be continuous. Assume 
\begin{align}
G(t, s, [\phi]) &= G(t, s, \phi(s), z(t, s, [\phi])) \\
z(t,s,[\phi]) &= \int_{s}^t k(t, t', s, \phi(t'))  dt'
\end{align}
to fullfill the following Lipschitz-type conditions: There are continuous functions {$L_1, L_2 \in C(D)$}, $D = \{(t, s) \in [t_0, \infty) \times [t_0, \infty) | s \leq t \}$, such that
\begin{align}
|G(t, s, x, z) - G(t, s, x', z')| \leq L_1(t, s) |x - x'| + L_2(t, s)|z - z'| 
\end{align}
and there is a continous function $L_k \in C(\tilde{D})$, \\ $\tilde{D} = \{(t, t', s)\in [t_0, \infty)^{3} | s\leq t' \leq t \}$, such that
\begin{align}
| k(t, t', s, x) - k(t, t', s, x')| \leq L_k(t, t', s) |x - x'| .
\end{align}
Assume furthermore, that $G(t, s, x, z)$ is bounded for $(t, s) \in D$ and $x, z \in \mathbb{R}^n$. 

Then the functional equation
\begin{align}
\eta u(t) = \sigma(t) - \int_{t_0}^t G\big(t, s, \phi(s), z(t, s, [\phi])\big) \,\, \phi(s) ds, \quad \eta > 0
\end{align}
has a unique (noncontinuable) solution $u(t)$, $t \in [t_0, T_{max})$, and if $u(t)$ stays bounded we have $T_{max} = \infty$.\footnote{Note, we chose $M = \mathbb{R}_{\geq t_0} \times \mathbb{R}^n$ for the formulation of this proposition}

\end{proposition}


\section{Convergence proof for collocation methods applied to Volterra equations with Fréchet-differentiable functional kernels} \label{app:convergence-proof}
\begin{theorem} \label{app:convergence-theorem}
Assume that the evolutionary integral equation
\begin{align}
u(t) = g(t) + \int_{t_0}^t K(t, s, [u]) ds, \quad t \in [t_0, T]
\end{align}
has a unique $m$-times continuously differentiable solution
$y \in C^{(m)}([t_0, T])$, $m \geq 1$, where $g(t)$ is the forcing function and the kernel 
$K(t, s, [u])$ is a (nonlinear) causal functional, e.g. $K(t, s, [u]) = K(t, s, \big[u|_{[t_0, t]}\big])$. 
Assume $K(t, s, [u])$ is continuously Fréchet-differentiable (in $u$) in $C([t_0, T])$ for all $s,t \in [t_0, T]$, $s \leq t$, 
and assume that the Fréchet-derivative $D_u K$ is of the form
\begin{align}
D_u K(t, s, [u]) \delta u = k_1(t, s, [u]) \delta u(s) + \int_{t_0}^t k_2(t, s, \tau, [u]) \delta u(\tau) d\tau, \quad \delta u \in C([t_0, T])
\end{align}
where $k_1$ and $k_2$ are bounded in the following sense:
\begin{align}
\sup \lVert k_1 \rVert_{L^1} &:= \sup_{u \in C([t_0,T]), t \in [t_0, T]} \int_{t_0}^t |k_1(t, s, [u])| ds < \infty
\\ \sup \lVert k_2 \rVert_{L^1} &:= \sup_{u \in C([t_0,T]), t \in [t_0, T]} \int_{t_0}^t \int_{t_0}^{t} |k_2(t, s, \tau, [u])| d\tau ds < \infty .
\end{align}
\\ Let $I_h = \{t_0, \dots, t_{N} \} \subset [t_0, T]$, $h = \max_{i} (t_{i+1} - t_i)$, be a mesh and let
\begin{align}
0 = c_1 < \dots < c_m = 1
\end{align}
be the set of collocation parameters. Then there exists a critical steplength $h_c$ such that 
\begin{align}
\lVert y(t) - u_h(t) \rVert_\infty \leq C \lVert y^{(m)}\rVert_\infty h^m 
\end{align}
holds for the collocation solution $u_h$ on any mesh with mesh diameter $h < h_c$.

If $u \mapsto K(t, s, [u])$ and $u \mapsto D_u K(t, s, [u])$ can be extended to piecewise continous and bounded functions $u \in PC_b([0, T])$ (with $L^\infty$-norm)
such that $\sup \lVert k_1 \rVert_{L^1} < \infty$ and $\sup \lVert k_2 \rVert_{L^1} < \infty$ also holds for $u \in PC_b([t_0, T])$, then the allowed range of collocation parameters can be extended to
\begin{align}
0 \leq c_1 < \dots < c_m \leq 1.
\end{align}
\end{theorem}
\begin{remark} Consider that the exact solution $y$ is known to lie in some open convex subset $F \subset C([t_0, T])$. This is for example the case if $y(t) > 0$ or if $|y(t)| < c$, $c \in \mathbb{R}_{>0}$, for all $t \in [t_0, T]$. Then, the full space $C([t_0, T])$ in the theorem above can be replaced by $F$, e.g. $K(t, s, [u])$ only needs to be Fréchet-differentiable in $F$ and the bounds $\sup \lVert k_1 \rVert_{L^1}$ and $\sup \lVert k_2 \rVert_{L^1}$ only need to be finite for $u \in F$. Note, the convexity of the subset $F$ is needed in the application of the vector-valued mean-value theorem in the proof and thus is a crucial condition.
\end{remark}

The derived order $h^m$ is a global convergence order. For collocation methods,
one often observes the phenomenon of local superconvergence, i.e., that
on the gridpoints the error converges faster than guaranteed by the
minimal convergence rate. 
We leave as an open question under which conditions local superconvergence can be guaranteed. Before we give the proof of the theorem, consider the following three applications:
\begin{example}[Nonlinear Volterra equations of second kind]
 In this case we have
\begin{align}
K(t, s, [u]) &= K(t, s, u(s))
\\ \Rightarrow D_u K(t, s, [u]) \delta u &= \partial_y K(t, s, y)|_{y = u(s)} \delta u(s)
\end{align}
Thus, according to the theorem convergence can be guaranteed if 
\begin{align}
\sup \lVert k_1 \rVert_{L^1} = \sup_{u \in C([t_0,T]), t \in [t_0, T]} \int_{t_0}^t |\partial_y K(t, s, u(s))| ds < \infty
\end{align}
This is for example the case if the partial derivative $\partial_y K(t, s, y)$ is bounded for all $y \in \mathbb{R}$ and $t, s \in [t_0, T]$ with $s < t$, see also \cite[Ch. 2.3.1 ]{Brunner2004} . 
In this case also local superconvergence has been shown \cite[Ch. 2.3.2]{Brunner2004} .
\end{example}
\begin{example}[Stressovershoot model] In this case the functional kernel is given by
\begin{align}
K(t, s, [u]) &= G(t, s, [u]) u(s)
\\G(t, s, [u]) &= G_\infty \bigg(1 - \big( \frac{\int_s^t u(\tau) d\tau }{\gamma_\ast}\big)^2 \bigg) e^{-\big(\int_{s}^t u(\tau) d\tau /\gamma_{\ast \ast} \big)^2} e^{-\frac{|u(s)|}{\gamma_c}}
\end{align}
Under the assumption, that $u$ does not change sign we can calculate the first Fréchet-derivative:
\begin{align}
D_u K(t, s, [u]) \delta u &= G(t, s, [u]) \delta u(s) + u(s) \,D_u G(t, s, [u]) \delta u
\\ D_u G(t, s, [u]) \delta u &= G_\infty \frac{-2}{\gamma_\ast} \big( \frac{\int_s^t u(\tau) d\tau }{\gamma_\ast}\big) e^{-\big(\int_{s}^t u(\tau) d\tau /\gamma_{\ast \ast} \big)^2} e^{-\frac{|u(s)|}{\gamma_c}} \int_{s}^t \delta u(\tau)\,  d\tau
\\& \quad +G_\infty \bigg(1 - \big( \frac{\int_s^t u(\tau) d\tau }{\gamma_\ast}\big)^2 \bigg) \bigg(\frac{-2}{\gamma_{\ast \ast}^2} \int_s^t u(\tau) \,d\tau \bigg) e^{-\big(\int_{s}^t u(\tau) d\tau /\gamma_{\ast \ast} \big)^2} e^{-\frac{|u(s)|}{\gamma_c}} \int_s^t \delta u(\tau)\, d\tau \nonumber
\\&\quad + G_\infty \bigg(1 - \big( \frac{\int_s^t u(\tau) d\tau }{\gamma_\ast}\big)^2 \bigg) e^{-\big(\int_{s}^t u(\tau) d\tau /\gamma_{\ast \ast} \big)^2} \frac{-1}{\gamma_c} e^{-\frac{|u(s)|}{\gamma_c}}  \,  \delta u(s)
\end{align}
Thus, the functions $k_1$ and $k_2$ from the theorem are given by
\begin{align}
k_1(t, s, [u]) &= G(t, s, [u]) - \frac{G(t, s, [u]) u(s)}{\gamma_c}
\\ k_2(t,s,\tau, [u]) &= \frac{-2}{\gamma_{\ast \ast}^2} \bigg(\int_s^t u(\tau) \,d\tau \bigg) G(t, s, [u]) u(s) \, \chi_{[s, t]}(\tau)
\\ &\quad - \frac{2 G_\infty}{\gamma_\ast} \big( \frac{\int_s^t u(\tau) d\tau }{\gamma_\ast}\big) e^{-\big(\int_{s}^t u(\tau) d\tau /\gamma_{\ast \ast} \big)^2} e^{-\frac{|u(s)|}{\gamma_c}} u(s) \, \chi_{[s, t]}(\tau) \nonumber
\end{align}
where $\chi_{[s, t]}(\tau)$ is the characteristic function, e.g.
$\chi_{[t, s]}(\tau) = 1$ if $\tau \in [s, t]$ and zero elsewhere.
Note, due to the exponential terms in $G(t, s, [u])$ both $k_1$ and $k_2$ are globally bounded and thus the collocation methods converge for the stressovershoot model. Since $G(t, s, [u])$ can be defined also for $u \in PC_b$ (and since $G(t, \cdot, [u]) \in L^\infty([t_0, t])$), collocation with discontinous splines is possible.
\end{example}
The additional assumptions placed on the first Frèchet derivative $D_u K$ to have a specific form may seem strong. However, note that $D_u K$ is a linear operator acting on the space of continous fuctions $C([t_0, T])$ (or on $PC_b([t_0, T])$ respectively). According to the theorem of Riesz and Markow (see for example \cite[Theorem 12.6]{Kubrusly2026}) 
any linear operator $A$ acting on continuous functions $\delta u \in C([t_0, T])$ can be represented by a measure $\nu$ such that
\begin{align}
A \delta u &= \int \delta u \,d\nu\,, &&\delta u \in C([t_0, T]) .
\end{align}
According to Lebesgue's decomposition theorem (see e.g. \cite[Corollary 4.17]{Kubrusly2026}), this measure can be decomposed into three parts: a part $\nu_c$ which can be described by a density $k$, a discrete part $\nu_d$ (point evaluations/Dirac-deltas $\delta(t_i - t)$) and a singular continuous part $\nu_{sc}$. Neglecting the singular continous part we set $\nu_{sc} = 0$, and we have
\begin{align}\nonumber
A \delta u &= \int \delta u \, d\nu_c +  \int \delta u \, d\nu_d\,, &&\delta u \in C([t_0, T])\,,
\\ &= \int_{[t_0, T]} k(t) \delta u(t) dt +  \sum_{i\in \mathbb{N}} c_i \delta u(t_i)\,,  && t_i \in [t_0, T].
\end{align}
Arguing that in physically motivated equations the point evaluations in $D_u K(t, s, [u]) \delta u$ should solely depend on $t$ and $s$ (and maybe $t_0$) and not on some other arbitrary time-instance, suggests to only consider point evaluations at these points\footnote{Including terms of the more general form $ \tilde{k}_i(t,s,[u]) \delta u\big( \theta_i(t, s) \big)$ would lead into the theory of Volterra equations with delays, which we will not discuss here.}. This implies to consider
\begin{align}
D_u K(t, s, [u]) \delta u = \int_{t_0}^t k_2(t, s,\tau, [u]) \delta u(\tau) \, d\tau + k_1(t, s, [u]) \, \delta u(s) 
\end{align}
Note, terms of the form $\tilde{k}_1(t, s, [u]) \, \delta u\big(t_0 + (t-s) \big)$ can be absorbed by the $k_1$-term using $\int_{t_0}^t \tilde{k}_1(t, s, [u]) \delta u\big(t_0+(t-s)\big) ds = \int_{t_0}^t \tilde{k}_1\big(t, t_0 + (t-s), [u]\big) \delta u(s) ds $ . 
\begin{example}[Generalized autoconvolution equation]
The generalized \acrfull{avie} is given by
\begin{align}
u(t) = g(t) + \int_{0}^t M(t, s) u(t-s) u(s) ds
\end{align}
It has been shown by \Citeauthor{Zhang.2016} that this equation has a unique solution $y \in C^{m}([0, T])$ if $g \in C^{m}([0, T])$ and $M \in C^{m}(D)$ 
\cite{Zhang.2016}. Furthermore, these authors also gave global convergence and local superconvergence results for collocation methods applied to the generalized \gls{avie}. 

Our theorem provides an alternative proof for the global convergence of collocation methods applied to the generalized \gls{avie} (but not for the local superconvergence): The functional kernel of the \gls{avie} is given by
\begin{align}
K(t, s, [u]) &= M(t, s) u(t-s) u(s)
\end{align}
Calculating the Fréchet-derivative $D_u K$ yields
\begin{align}
D_u K(t, s, [u]) \delta u &= M(t, s) u(s) \delta u(t-s) + M(t, s) u(t-s) \delta u(s) \quad , \delta u \in C([0, T])
\end{align}
Thus, the corresponding $k_1(t, s, [u])$ from the theorem is given by
\begin{align}
k_1(t, s, [u]) = M(t, t-s) u(t-s) + M(t, s) u(t-s) 
\end{align}
Since we know a solution $y(t)$ exists in $C^m([0, T])$, $|y(t)|$ assumes a maximum $y_{max}$ somewhere. Thus, there is a constant $c > y_{max}$. From now on only consider functions $u \in F_c = \{f \in C([0, T]); |f(t)| < c\}$. Then $y \in F_c$ and
\begin{align}
\sup \lVert k_1 \rVert_{L^1} &= \sup_{u \in F_c, t \in [t_0, T]} \int_{t_0}^t | k_1(t, s, [u]) | ds
\\&\leq c T  \max_{0 \leq s \leq t \leq T} |M(t, s)| < \infty
\end{align} 
Applying theorem \ref{app:convergence-theorem} and the remark after it then yields the same global convergence order as also has been found by \Citeauthor{Zhang.2016} \cite{Zhang.2016}. 
\end{example}

\begin{proof}[Proof of Theorem \ref{app:convergence-theorem}]
We adapt the proof given by \citeauthor{Brunner2004} for nonlinear Volterra integral equations \cite[Ch. 2.3.1]{Brunner2004}, generalizing the underlying idea to the case of functional kernels $K(t, s, [u])$. 
The underlying idea is to linearize the kernel using a mean-value theorem and then proceed as in the linear case by deriving a discrete Gronwall inequality (compare theorem \ref{app:discrete-Gronwall} below) which will establish an upper bound of order ${O}(h^m)$ in combination with some kind of boundedness properties of the Frèchet-derivative $D_u K$.

Assume the problem has a unique solution $y(t)$ with $y \in C^{m}[0, T]$. 
Let $t_0 < \dots < t_N = T$ be the gridpoints and denote the collocation points by
$t_{n,k} = t_n + c_k h_n$
with the collocation parameters $0 \leq c_1 < \dots < c_m \leq 1$. If the functional $K(t, s, [u])$ cannot be extended to piecewise continuous and bounded functions $u \in PC_b([t_0, T])$, set $c_1 = 0$ and $c_m = 1$.
We denote the collocation solution by
\begin{align}
u(t) &= \sum_{n = 0}^{N-1} \chi_{(t_n, t_{n+1}]}(t) \sum_{k = 1}^m u(t_{n,k}) L_k\bigg(\frac{t - t_n}{h_n}\bigg)
\end{align}
where $L_k$ are the Lagrange-Polynomials corresponding to the abscissas $c_k$ on the intervall $[0, 1]$. Note, in the case $c_1 = 0$ and $c_m = 1$ the collocation solution $u(t)$ is continuous by construction such that $K(t, s, [u])$ is well-defined even if $K(t, s, [u])$ cannot be extended to $u \in PC_b$. \\
It will be convenient to represent $u(t)$ as a sum of some kind of base functions.
We choose\footnote{In the discontinous case this indeed a basis of the space of discontinous splines $p \in S^{(-1)}_m(I_h)$. In the continous case ($p \in S^{(0)}_m(I_h)$) this does strictly speaking not define a basis, since the $\tilde{L}_{i, 1}$ and $\tilde{L}_{i, m}$ are discontinous. However, this does not pose a problem for this proof.}
\begin{align}
\tilde{L}_{i, j}(t) &= \chi_{(t_{i}, t_{i+1}]}(t) L_j(\frac{t - t_{i}}{h_{i}}), \quad i = 0, \dots, N-1; j = 1, \dots, m
\end{align}
Thus we have for $t \in [t_0, t_N]$
\begin{align}
u(t) &= \sum_{n = 0}^{N-1} \sum_{j = 1}^m U_{n,j} \tilde{L}_{n, j}(t) 
\end{align}

Our first task will be to show that the collocation equation actually has a (unique) solution. Then we will show in several steps that this solution indeed converges globally(!) to the exact solution $y(t)$ as the grid gets more refined ($h \rightarrow 0$). 

\newcounter{stepcounter}
\setcounter{stepcounter}{1}
\quad \textit{Step \thestepcounter : The collocation equation has a solution for $h \leq h_c$}\\
The collocation equations are given by
\begin{align}
U_{n,k} = u(t_{n,k}) = g(t_{n,k}) + \int_{t_0}^{t_{n,k}} K(t_{n,k}, s, [u]) ds
\end{align}
Assume we already calculated the values $U_{i,j}$ for $i = 0, \dots, n-1$ and $j = 1, \dots ,m$. The right hand side of the collocation equation then defines a function
\begin{align}
&f: \mathbb{R}^m \rightarrow \mathbb{R}^m
\\ & \begin{pmatrix} U_{n,1} \\ \vdots \\ U_{n, m} \end{pmatrix} 
\mapsto 
\begin{pmatrix} 
g(t_{n,1}) + \int_{t_0}^{t_{n,1}} K(t_{n,m}, s, [u]) ds \\ \vdots \\ g(t_{n,m}) + \int_{t_0}^{t_{n,m}} K(t_{n,m}, s, [u]) ds
\end{pmatrix}
\\ &u(t) = \sum_{j = 1}^m U_{n,j} \tilde{L}_{n, j}(t) + \sum_{i = 0}^{n-1} \sum_{j = 1}^m U_{i,j} \tilde{L}_{i, j}(t)
\end{align}
Note, the fixpoints of $f$ are the solutions to the collocation equations. We will now use Banachs fixpoint theorem to show that $f$ indeed has a unique fixpoint as long as $h_n$ is smaller than some critical steplength $h_c$. Let $U, V \in \mathbb{R}^m$ be two vectors. Then
\begin{align}
\big(f(U) - f(V)\big)_k = \int_{t_0}^{t_{n,k}} \big(K(t_{n,k}, s, [u_1]) - K(t_{n,k}, s, [u_2]) \big) ds
\end{align}
with 
\begin{align}
u_1(t) &= \sum_{j = 1}^m U_{j} \tilde{L}_{n, j}(t) + u_{\text{past}}(t)
\\ u_2(t) &= \sum_{j = 1}^m V_{j} \tilde{L}_{n, j}(t) + u_{\text{past}}(t)
\end{align}
We can linearize the equation by employing the vector-valued mean-value theorem (see theorem  \ref{app:mean-value-theorem} below)
\begin{align}
& K(t_{n,k}, s, [u_1]) - K(t_{n,k}, s, [u_2]) = \int_0^1 D_u K(t_{n,k}, s, [u_2 + \alpha (u_1 - u_2)]) (u_1 - u_2) d\alpha 
\\ & u_1(t) - u_2(t) = \sum_{j = 1}^m (U_j - V_{j}) \tilde{L}_{n, j}(t)
\end{align}
Inserting our assumed form of the Fréchet-derivative 
\begin{align}
|\big(f(U) - f(V)\big)_k| &\leq \bigg| \int_{t_0}^{t_{n,k}} \int_0^1 k_1(t_{n,k}, s, [u_1 + \alpha (u_2 - u_1) ]) (u_1(s) - u_2(s)) d\alpha ds \bigg|
\\& \quad + \bigg| \int_{t_0}^{t_{n,k}} \int_0^1 \int_{t_0}^t k_2(t_{n,k}, s, \tau, [u_1 + \alpha (u_2 - u_1) ]) (u_1(\tau) - u_2(\tau)) d\tau d\alpha ds \bigg|
\\&= \bigg| h_n \sum_{j = 1}^m (U_j - V_j) \int_{0}^{c_k} \int_0^1 k_1(t_{n,k}, t_n+s h_n, [u_1 + \alpha (u_2 - u_1) ]) L_j(s) d\alpha ds \bigg|
\\& \quad + \bigg| h_n \sum_{j = 1}^m (U_j - V_j) \int_{t_0}^{t_{n,k}} \int_0^1 \int_{0}^1 k_2(t_{n,k}, s, t_n + \tau h_n, [u_1 + \alpha (u_2 - u_1) ]) L_j(\tau) d\tau d\alpha ds \bigg|
\\&\leq h_n C_k \big( \sup \lVert k_1 \rVert_{L^1} + \sup \lVert k_2\rVert_{L^1}\big) \lVert U - V \rVert
\end{align}
From this follows, that for $h < h_c$ $f$ is a contraction and thus the collocation equations have unique solutions $U_{n, k}$, $k = 1, \dots, m$. 

\addtocounter{stepcounter}{1}
\quad \textit{Step \thestepcounter: Identifying the collocation error}\\
The collocation error is defined as
\begin{align}
e(t) = y(t) - u(t)
\end{align}
To get a better representation, recall Peano's theorem for $t \in [t_n, t_{n+1}]$ for the error of polynomial interpolation (theorem \ref{app:peanos-theorem} below):
\begin{align}
y(t_n + z h_n) &= \sum_{k = 1}^m y(t_{n,k}) L_j(z) + h_n^{m} \int_{0}^{1} K_m(z, \tau) f^{(m)}(t_n + \tau h_n) d\tau 
\\ &= \sum_{k = 1}^m y(t_{n,k}) L_k(z) + h_n^{m} \rho_n(z)
\end{align}
Writing $Y_{n, k} = y(t_{n,k})$ we have for $t \in (t_0, T]$:
\begin{align}
y(t) &= \sum_{n = 0}^{N-1} \chi_{(t_n, t_{n+1}]} \sum_{k = 1}^m y(t_{n, k}) L_k\bigg(\frac{t - t_n}{h_n}\bigg) + \sum_{n = 0}^N \chi_{(t_n, t_{n+1}]} h_n^{m} \rho_n\bigg( \frac{t - t_n}{h_n} \bigg)
\\ &= \sum_{n = 0}^{N-1} \sum_{k = 1}^m Y_{n, k} \tilde{L}_{n, k} + \sum_{n = 0}^N h_n^m \tilde{\rho}_m(t)
\end{align}
where we defined 
\begin{align}
\tilde{\rho}_n(t) = \chi_{(t_n, t_{n+1}]}(t) \rho_n\bigg(\frac{t-t_n}{h_n} \bigg) . \label{equ_conv:error-polyint-1}
\end{align}
Thus, the collocation error is given by
\begin{align}
e(t) &= \sum_{n = 0}^{N-1} \sum_{k = 1}^m E_{n,k} \tilde{L}_{n,k}(t) + \sum_{n = 0}^{N-1}  h_n^{m} \tilde{\rho}_n(t) \label{conv:error-decomposition}
\\ E_{n,k} &= Y_{n,k} - U_{n,k}
\end{align}
To show convergence, we thus have to show on one hand that the $\rho_n(t)$ are bounded by a common constant and that the error $E_{n,k} = e(t_{n,k})$ at the collocation point scales with the maximum steplength $h$. 
As Peano's theorem shows, the interpolation error $\rho_n$ will be bounded if $\lVert y^{(m)} \rVert_\infty < \infty$. We will deal with $E_{n,k}$ by using a Gronwall inequality.

\addtocounter{stepcounter}{1}
\quad \textit{Step \thestepcounter: Deriving an inequality for the evolution of the collocation error} \\
Since the collocation solution solves the equation exactly at the collocation points and since the interpolation error $\rho_n(t_{n,k}) = 0$ vanishes on the collocation points, the collocation error solves the following equation:
\begin{align}
e(t_{n,k}) = E_{n,k} &= \int_{t_0}^{t_{n,k}} \bigg( K(t_{n,k}, s, [y]) - K(t_{n,k}, s, [u])\bigg) ds
\end{align}
Again, we can 'linearize' using the mean-value theorem (theorem \ref{app:mean-value-theorem}):
\begin{align}
E_{n,k} &= \int_{t_0}^{t_{n,k}} \bigg( \int_{0}^1 D_u K(t_{n,k}, s,[u + \alpha (y-u)]) \,(y - u) d\alpha \bigg) ds
\\ &= \int_{t_0}^{t_{n,k}} \bigg( \int_{0}^1 D_u K(t_{n,k}, s,[u + \alpha e]) \,e\, d\alpha \bigg) ds
\end{align}
Next, insert the representation of the collocation error, Eq.~\ref{conv:error-decomposition}),
\begin{align}
E_{n, k} &= 
\sum_{j = 1}^m E_{n,j} \int_{t_0}^{t_{n,k}} \bigg( \int_{0}^1 D_u K(t_{n,k}, s,[u + \alpha e]) \,\big( \tilde{L}_{n,j} \big) d\alpha \bigg) ds
\\ &+ \sum_{i = 0}^{n-1} \sum_{j = 1}^{m} E_{i,j} \int_{t_0}^{t_{n,k}} \bigg( \int_{0}^1 D_u K(t_{n,k}, s,[u + \alpha e]) \,\big( \tilde{L}_{i,j} \big) d\alpha \bigg) ds
\\ &+ \sum_{i = 0}^{n} h_i^m \int_{t_0}^{t_{n,k}} \bigg( \int_{0}^1 D_u K(t_{n,k}, s,[u + \alpha e]) \,\big( \tilde{\rho}_i \big) d\alpha \bigg) ds
\end{align}
Here, we already used the causality of the operator $K$ to rule out contributions from future times $t > t_{n,k}$. 
Defining the vector $E_n = (E_{n,1}, \dots, E_{n, m})$ we have a linear equation for $E_n$ of the form
\begin{align}
\big(\mathds{1} - B_n \big) E_n = \sum_{i = 0}^{n-1} B_{i} E_{i} + \sum_{i = 0}^n h_i^m P_i
\end{align}
Assuming that $(\mathds{1} - B_n)$ is invertible and uniformly bounded for all $n \geq 0$ ($\lVert (\mathds{1} - B_n)^{-1} \rVert \leq D_0$), we have the following inequality for the norm of $E_n$:
\begin{align}
\lVert E_0 \rVert &\leq h_0^m \lVert (\mathds{1} - B_0)^{-1} \rVert \, \lVert P_0 \rVert \leq h_0^m D_0 \lVert P_0 \rVert \label{equ:proof-E0-bound1}
\\
\lVert E_n \rVert &\leq  \sum_{i = 0}^{n-1}  \lVert (\mathds{1} - B_n)^{-1} \rVert\, \lVert B_{i} \rVert \, \lVert E_{i} \rVert + \sum_{i = 0}^n h_i^m \lVert (\mathds{1} - B_n)^{-1} \rVert \lVert P_i \rVert, \quad n \geq 1
\\
&\leq \sum_{i = 0}^{n-1} D_0 \lVert B_{i} \rVert \, \lVert E_{i} \rVert + \sum_{i = 0}^n h_i^m D_0 \lVert P_i \rVert, \quad n \geq 1 \label{equ:proof-En-Gronwall}
\end{align}
Next, we apply Gronwalls inequality (theorem \ref{app:discrete-Gronwall}) to the sequence $\epsilon_n = \lVert E_n \rVert$, $k_j = D_0 \lVert B_j \rVert$, $q_j = h_{j+1}^m D_0 \lVert P_{j+1} \rVert$, $\rho_0 = h_0^m D_0 \lVert P_0 \rVert$, and get the inequality
\begin{align}
\lVert E_n \rVert \leq \bigg( D_0 \sum_{i = 0}^{n} h_i^m \lVert P_{i} \rVert \bigg) \exp\bigg( D_0 \sum_{i = 0}^{n-1}  \lVert B_{i} \rVert \bigg)
\end{align}
To establish convergence we now have to show that the terms appearing on the right hand side are bounded and vanish as $h \rightarrow 0$ as will be done in the next few steps of the proof.

\addtocounter{stepcounter}{1}
\quad \textit{Step \thestepcounter: $(\mathds{1}-B_n)$ is invertible and the inverse stays bounded as $h \rightarrow 0$} \\
The idea is to use the Neumann-series to establish the existence of an inverse. 
Thus, we need to show $\rVert B_n \lVert \rightarrow 0$ as $h \rightarrow 0$, since then there is a critical steplength $h_c$ such that $(\mathds{1} - B_n)$ is invertible and the inverse is bounded by $\lVert (\mathds{1} - B_n)^{-1} \rVert \leq D_0$ for all $h \leq h_c$ for a $D_0 \in \mathbb{R}_{+}$. 
\\ 
The matrix $B_n$ is defined by
\begin{align}
(B_n)_{k, j} = \int_{t_0}^{t_{n,k}} \bigg( \int_{0}^1 D_u K(t_{n,k}, s,[u + \alpha e]) \,\big( \tilde{L}_{n,j} \big) d\alpha \bigg) ds
\end{align}
Inserting the assumed form of the Frechet-derivative $D_u K$ yields
\begin{align}
(B_n)_{k,j} &= \int_{t_0}^{t_{n,k}} \bigg( \int_{0}^1 k_1(t_{n,k}, s, [u + \alpha e]) d\alpha \bigg) \tilde{L}_{n,j}(s) ds 
\\ &\quad + \int_{t_0}^{t_{n,k}} \bigg( \int_{0}^1 \big( \int_{t_0}^{t_{n,k}} k_2(t_{n,k}, s, \tau, [u + \alpha e]) \tilde{L}_{n,j}(\tau) d\tau \big) d\alpha \bigg) ds
\end{align}
Restricting the integrals to the support of $\tilde{L}_{j,k}$ and applying the transformation $t \rightarrow \frac{t - t_j}{h_j}$ yields:
\begin{align}
(B_n)_{k,j} &= h_n \int_{0}^{c_k} \bigg( \int_0^1 k_1(t_{n,k}, t_n + s h_n, [u + \alpha e]) d\alpha \bigg)  L_j(s) ds
\\&\quad + h_n \int_{t_0}^{t_{n,k}} \int_0^1 \bigg( \int_{0}^{c_k} k_2(t_{n,k}, s, t_n + \tau h_n, [u + \alpha e]) L_j(\tau) d\tau \bigg) d\alpha ds
\end{align}
Since $\sup |k_1|, \sup |k_2| < \infty$ are bounded, we have
\begin{align}
|(B_n)_{k,j}| &\leq h_n \sup \lVert k_1 \rVert_{L^1}  \max_{s \in [0,1]} |L_j(s)| + h_n \sup \lVert k_2 \rVert_{L^1} \max_{s \in [0,1]} |L_j(s)|
\\ &\leq h_n C_j \big( \sup \lVert k_1 \rVert_{L^1}  +  \sup \lVert k_2 \rVert_{L^1} \big)  
\end{align} 
Thus, all elements of $B_n$ are bounded by a factor proportional to $h_n$ and there exists an $h_c$ such that $(\mathds{1} - B_n)$ is invertible with bounded inverse for all $h < h_c$.

\addtocounter{stepcounter}{1}
\quad \textit{Step \thestepcounter: Bounds for $\lVert B_i \rVert$} \\
The elements of the matrices $B_i$, $i = 0, \dots, n-1$ were defined by
\begin{align}
(B_i)_{k, j} &= \int_{t_0}^{t_{n,k}} \bigg( \int_0^1 D_u K(t_{n,k}, s, [y + \alpha e])(\tilde{L}_{i, j})  d\alpha \bigg) ds
\\ &= h_i \int_{0}^1 \bigg(\int_0^1 k_1(t_{n,k}, t_i + s h_i, [y+\alpha e]) d\alpha \bigg) L_j(s) ds
\\ &\quad+ h_i \int_{t_0}^{t_{n,k}} \int_0^1 \bigg( \int_{0}^{1} k_2\big(t_{n,k}, s, t_i + \tau h_i, [y + \alpha e] \big) L_j(\tau) d\tau \bigg) d\alpha ds
\end{align}
Thus, we have
\begin{align}
|(B_i)_{j,k}| \leq h_i C_j \bigg( \sup \lVert k_1 \rVert_{L^1} + \sup \lVert k_2 \rVert_{L^1} \bigg)
\end{align}

\addtocounter{stepcounter}{1}
\quad \textit{Step \thestepcounter: Bounds for $\lVert P_i \rVert$} \\
The elements of $P_i$ are given by
\begin{align}
(P_i)_k &= \int_{t_0}^{t_{n,k}} \bigg( \int_0^1 D_u K(t_{n,k}, s, [u + \alpha e]) \tilde{\rho}_i d\alpha \bigg) ds
\\&= \int_{t_0}^{t_{n,k}} \bigg( \int_0^1 k_1(t_{n,k}, s, [u + \alpha e]) d\alpha \bigg) \tilde{\rho}_i(s) ds
\\&\quad+\int_{t_0}^{t_{n,k}} \int_0^1 \bigg( \int_{t_0}^{t_{n,k}} k_2(t_{n,k}, s, \tau, [u + \alpha e]) \tilde{\rho}_i(\tau) d\tau \bigg) d\alpha  ds \nonumber
\end{align}
Inserting the definition of $\tilde{\rho}_i$ (equ.  \ref{equ_conv:error-polyint-1} ) yields
\begin{align}
(P_i)_k &= h_i \int_{0}^{1} \bigg( \int_0^1 k_1(t_{n,k}, t_i + s h_i, [u + \alpha e]) d\alpha \bigg) \rho_i(s) ds
\\&\quad+ h_i \int_{t_0}^{t_{n,k}} \int_0^1 \bigg( \int_{0}^{1} k_2(t_{n,k}, s, t_i + \tau h_i, [u + \alpha e]) \rho_i(\tau) d\tau \bigg) d\alpha  ds \nonumber
\end{align}
Thus, we have
\begin{align}
|(P_i)_k| \leq h_i \bigg(\sup \lVert k_1 \rVert_{L^1} + \sup \lVert k_2 \rVert_{L^1} \bigg) \max_{\tau \in [0, 1]} |\rho_i(\tau)| 
\end{align}
Recall from Peano's theorem (theorem \ref{app:peanos-theorem}) that $\rho_n(\tau) = \int_0^1 K_m(z, \tau) f^{(m)}(t_n + z h_n) dz$. This yields
\begin{align}
|(P_i)_k| \leq h_i C_m \bigg(\sup \lVert k_1 \rVert_{L^1} + \sup \lVert k_2 \rVert_{L^1} \bigg) \lVert y^{(m)}\rVert_\infty 
\end{align} 

\addtocounter{stepcounter}{1}
\quad \textit{Step \thestepcounter: Bound for $\lVert E_0 \rVert$} \\
As previously derived when establishing the Gronwall inequality (equ. \ref{equ:proof-E0-bound1}), we have
\begin{align}
\lVert E_0 \rVert \leq h_0^m D_0 \, \lVert P_0 \rVert
\end{align}
Inserting the bounds for the $| (P_0)_k |$ derived in the previous step yields
\begin{align}
\lVert E_0 \rVert \leq h_0^{m+1} \tilde{C}_m \bigg(\sup \lVert k_1 \rVert_{L^1} + \sup \lVert k_2 \rVert_{L^1} \bigg) \lVert y^{(m)}\rVert_\infty
\end{align}
The exact definition of $\tilde{C}_m$ now depends on the choice of the vector norm $\lVert E_0 \rVert$. If the $1$-norm has been choosen we have $\tilde{C}_m  = m \,D_0 C_m$.

\quad \textit{Last step: Inserting all previous results into the discrete Gronwall Inequality} \\
Inserting the previous bounds into the Gronwall inequality, Eq.~\eqref{equ:proof-En-Gronwall}, yields
\begin{align}
\lVert E_n \rVert &\leq \bigg( \sum_{i = 0}^{n} h_i^m \tilde{C}_m h_i \lVert y^{(m)}\rVert_\infty \big(\sup \lVert k_1 \rVert_{L^1} + \sup \lVert k_2 \rVert_{L^1} \big) \bigg) e^{\sum_{i = 0}^{n} h_i \bar{C} \big(\sup \lVert k_1 \rVert_{L^1} + \sup \lVert k_2\rVert_{L^1} \big)}
\\ &\leq h^m T \lVert y^{(m)}\rVert_\infty \big(\sup \lVert k_1 \rVert_{L^1} + \sup \lVert k_2 \rVert_{L^1} \big) e^{ \bar{C} T \big(\sup \lVert k_1 \rVert_{L^1} + \sup \lVert k_2\rVert_{L^1} \big) }
\\ &\leq C(T) \lVert y^{(m)} \rVert_\infty h^m
\end{align}
This prooves the convergence theorem. 
\end{proof}

\subsection{Collection of theorems from Analysis needed for the proof}\label{app:analysis-theorems}

\begin{theorem}[Mean-Value theorem] \cite[Ch.VII Theorem 3.10]{Amann2008}\label{app:mean-value-theorem}
\\ Let $f \in C^1(X, F)$ be a continuous (Fréchet-)differentiable function between two Banach-spaces $E, F$ with $X\subseteq E$. Then:
\begin{align}
f(y) - f(x) = \int_0^1 D f\big(x + \alpha(y-x)\big) \, (y -x)\,d\alpha
\end{align}
as long as $[x, y] = \{x + \alpha(y-x) | \alpha \in [0,1]\} \subset X$. 
\end{theorem}

\begin{theorem}[Peano's Theorem] \cite[Theorem 1.8.1 and corollary 1.8.2]{Brunner2004}\label{app:peanos-theorem}
\\For given abscissas $a \leq \xi_1 < \dots < \xi_m \leq b$ with $\xi_i = a + c_i h$, $h = b-a$,   let 
\begin{align}
e_m(f;t) = f(t) - \sum_{j = 1}^m L_j\bigg(\frac{t-a}{h}\bigg) f(\xi_j), \quad t \in [a, b]
\end{align}
the error of the polynomial interpolation, with $L_j$ the Lagrange-polynomials on the intervall $[0,1]$ with abscissas $c_j$. Assume that $f \in C^d([a, b])$ is $d$-times continuously differentiable with $1 \leq d \leq m$. Then the interpolation error $e_m(f; t)$ posseses the integral representation
\begin{align}
e_m(f; a + z h) = h^d \int_0^1 K_d(z, s) f^{(d)}(a + s h) ds
\end{align}
where the Peano-kernel $K_d$ is given by
\begin{align}
K_d(z, s) &= \frac{1}{(d-1)!} \bigg( (z-s)^{d-1}_{+} - \sum_{k = 1}^m L_k(z) (c_k - s)^{d-1}_{+} \bigg)
\\ (z-s)^p_{+} &= \begin{cases}  0 &,\text{ if } z < s \\ (z-s)^p &,\text{ if } z \geq s  \end{cases}
\end{align} 
\end{theorem}

\begin{theorem}[Discrete Gronwall Inequality] \cite[Theorem 2.1.18]{Brunner2004}\label{app:discrete-Gronwall}
\\Assume that $\{k_j\}_{j\geq 0}$ is a given non-negative sequence ($k_j \geq 0$). If a sequence $\{\epsilon_n\}_{n\geq 0}$ saitsfies
\begin{enumerate}
\item) $\epsilon_0 \leq \rho_0$
\item) $\epsilon_n \leq \rho_0 + \sum_{j = 0}^{n-1} q_j + \sum_{j = 0}^{n-1} k_j \epsilon_j $ 
\end{enumerate}
with $\rho_0 \geq 0$, $q_j \geq 0$ for $j \geq 0$, then it fulfills the discrete Gronwall inequality:
\begin{align}
\epsilon_n \leq \bigg(\rho_0 + \sum_{j = 0}^{n-1} q_j \bigg) e^{\sum_{j = 0}^{n-1} k_j}
\end{align}
\end{theorem}

\section{Trapezoidal rule applied to functions with jump discontinuities and kinks} \label{appendix:trapezoidal_jumps}
We want to understand how isolated jump-discontinuities and kinks in the integrand $f(t)$ influence the convergence behaviour of the (iterated) trapezoidal rule for numerical integration of $\int_a^b f(t) dt$.  
Amat et al. \cite[Theorem 1]{Amat2023} showed, that the error of the trapezoidal rule depends on the jump terms
\begin{align}
[f](t^\ast) &= f(t^{\ast}_{+}) - f(t^{\ast}_{-})
\\ [f'](t^\ast) &= f'(t^{\ast}_{+}) - f'(t^{\ast}_{-})
\end{align}
where $t^\ast$ denotes the position of the jump discontinuity or kink of $f$. Let
\begin{align}
a = t_0 < \dots < t_n = b 
\end{align}
denote the grid used for the numerical integration. The local steplength is denoted as
\begin{align}
h_j &= t_{j} - t_{j-1} \quad j \in \{0, \dots, n-1 \} 
\end{align}
Let $t_k^\ast \in [a, b]$, $k \in \{1, \dots, m\}$, denote the position of the (isolated) jumps or kinks of $f$ in $[a, b]$. Denote the intervall where the jumps or kinks are located by $t_k^\ast \in [t_{j_{k}}, t_{j_{k} + 1}]$ and assume that each such interval contains only one such point. The distance from the jump or kink to the lower endpoint of the interval is given by
\begin{align}
\alpha_k &= t_k^\ast - t_{j_k} \quad k \in \{1, \dots, m\} .
\end{align}
The local error of the trapezoidal approximation given in the paper from Amat et al. \cite{Amat2023} then reads
\begin{align}
\int_{t_{j_k}}^{t_{j_k + 1}} f(t) dt - \frac{h_{j_k}}{2} \big( f(t_{j_k+1}) + f(t_{j_k}) \big) =& \frac{h_{j_k} -  2 \alpha_k}{2} [f](t_k^\ast) + \frac{\alpha_k^2 - h_{j_k} \alpha_k}{2} [f'](t_k^\ast)
\\ & - \frac{1}{12} \alpha_k^3  f''(\xi_1) - \frac{1}{12}   (h_{j_k} - \alpha_k)^3 f''(\xi_2)  + O(h_j^4) \nonumber
\end{align}
where $\xi_1 \in [t_{j_k}, t^\ast_k]$ and $\xi_2 \in [t^\ast_k, t_{j_k + 1}]$. 
Thus, if the positions of the kink $t_k^\ast$ and the jump terms $[f](t_k^\ast)$ and $[f'](t_k^\ast)$ are known, one can calculate corrections to the interated trapezoidal rule to ensure that the (global) approximation
can be bounded by $O(h_{max}^2)$ for a sequence of quasiuniform grids\footnote{A sequence of quasiuniform grids is such that $h_{max}/h_{min} \leq C$, $1 \leq C < \infty $ is true for all grids in the sequence. In the case $C = 1$ one has a sequence of equidistant grids ($h_j = h_{max} = h_{min} = h$)}:
\begin{align}
&\bigg| \int_a^b f(t) dt - \sum_{j = 0}^{n - 1} \frac{h_j}{2} (f(t_{j+1}) + f(t_j)) + \sum_{k = 1}^m C_k + C'_k \bigg| \leq \frac{(b-a)}{12} \frac{h_{max}}{h_{min}} h_{max}^2   \max_{\xi \in [a, b]} f''(\xi) \label{equ:app-num-1}
\\ &C_k = \frac{h_{j_k} -  2 \alpha_k}{2} [f](t_k^\ast)
\\ &C'_k = \frac{\alpha_k^2 - h_{j_k} \alpha_k}{2} [f'](t_k^\ast)
\end{align}  
Note, during refinement of the grid $\{t_j\}$ the distances
\begin{align}
\alpha_k = t_k^\ast - t_{j_k} \leq h_{j_k}
\end{align}
between the kink positions $t_k^\ast$ and the nearest smaller gridpoints $t_{j_k}$ fluctuate arbitrarily in general.Thus the error will fluctuate too and even with the correction one will only see a 'rugged' second order convergence which prohibits the use of extrapolation methods. 
\\ To apply this result to the numerical study of the SRR and the SO model it is necessary to look at the jump terms $[G(t, s, [u])] u(s)$ and $[\partial_s \big( G(t, s, [u]) u(s) \big) ]$ at the points $t^\ast$ where $u(t^\ast) = 0$. 
Since $G(t, s, [u])$ is continuous and smooth everywhere where $u(t)$ does not vanish, we have
\begin{align}
[G(t, t, [u]) u(t)] = 0
\end{align}
To calculate the jump in the first derivative note
\begin{align}
\partial_s \big( G(t, s, [u]) u(s) \big) =\quad& G(t, s, [u]) \partial_s u(s) + \big(\partial_s G(t, s, [u]) \big) u(s) 
\\ \stackrel{u(s) = 0}{=}& G(t, s, [u]) \partial_s u(s)
\end{align}
Thus, for the SRR and the SO model we have
\begin{align}
[ \partial_{s = t^\ast}\big( G(t, t^\ast, [u]) u(s) \big) ] = 0
\end{align}
Therefore, for both models the approximation to the integral term using the iterated trapezoidal rule is of second order by equation \ref{equ:app-num-1} even if $u(t)$ changes sign, e.g. if $u(t^\ast) = 0$ at isolated points $t^\ast$. 

\end{appendix}

\end{document}